# Predicting Resistive Pulse Signatures in Nanopores by Accurately Modeling Access Regions

Martin Charron, Zachary Roelen, Deekshant Wadhwa, Vincent Tabard-Cossa*

150 Louis-Pasteur Private, Department of Physics, University of Ottawa, Ottawa K1N 6N5, Canada

*Corresponding Author: tcossa@uottawa.ca

**Abstract**

Resistive pulse sensing is used to characterize and count single particles in solution moving through channels under an electric bias, with nanoscale pores providing enough spatial resolution for single-molecule identification and sequencing. This technique relies on measuring the ionic current drop produced by the passage of a molecule and, through conductance models, translating the blockage signal into molecular dimensions. However, no generalized model exists that considers the resistive contributions of the pore exterior, i.e. the access regions, when obstructed by a molecule. This is becoming increasingly important with the advent of 2D materials and ultrathin membranes featuring low aspect ratio pores. In this work, a general method by which to model the resistance of access regions in the presence of an insulating obstruction is presented. Thin oblate spheroidal slices are used to partition access regions and infer their conductance when blocked by differently shaped objects. We show that our model accurately estimates the blocked-state conductance of 2D and finite-length pores in the presence of simple or complex structures positioned at varying distances from the pore opening. The model is shown to capture off-axis effects by predicting deeper blockages for obstructions offset from the pore's central axis. A model-based web app was created to predict the electrical signatures of a wide range of molecule geometries translocating through differently shaped pores. The introduced model and accompanying tool will help guide future experimental designs and thus present a straightforward way to extend the quantification of the resistive pulse technique at the nanoscale.

Resistive pulse sensing is a well-established single-particle detection method that relies on measuring the reduction of current induced by the passage of an object through a fluid-embedded channel across which a potential difference is applied (Fig. 1).[1–6] Since the current blockage amplitude is closely related to the volume of the current-obstructing object, microchannels have been used to count and size cells, viruses and colloidal particles.[1,2] More recently, nanopores have pushed the resolution of this technique further by sensing single molecules of DNA and proteins as well as other nanostructures for different sequencing, diagnostics or next-generation information storage applications.[3–8]

At the heart of nanopore sensing and more generally the resistive pulse method, mathematical models are required to transduce blockage amplitudes into molecular dimensions so as to characterize and identify translocating objects.[9,10] For cylindrical pores whose length $L_p$ is much larger than their diameter $d_p$, the potential drop $\Delta V$ occurs almost entirely inside the channel, and as such the electric field strength inside the pore is estimated to be constant $E = \Delta V/L_p$. Under these conditions, the interior of the channel can be partitioned into circular slices of thickness $dz$, and an accurate estimate of the channel resistance can be obtained by summing up the $z$-dependent resistance $dR(z)$ of each slice:

$$R = \int dR\,(z) = \frac{1}{\sigma} \int_{-\frac{L_p}{2}}^{\frac{L_p}{2}} \frac{dz}{A(z)} \tag{1}$$

Here, $\sigma$ and $A(z)$ denote the bulk conductivity and the $z$-dependent conductive area of each circular slice. Equation 1 can either be applied to open pores, or to pores blocked by an insulating obstruction.[11] For example, a linear polymer with cylindrical cross-section of radius $r_c$ passing through a cylindrical pore of radius $r_p$, simply results in $A(z) = \pi(r_p^2 - r_c)$. Equation 1 is generalizable to non-cylindrical obstructions and pores, and can be improved upon for specific use cases using empirical correction factors.[9,12–15] An important result from Eq. 1 is that the resistance increase $\Delta R$ upon the introduction a molecule of volume $\mathcal{V}_{mol}$ inside a pore channel of volume $\mathcal{V}_{pore}$ with an open pore resistance of $R_o$ scales as $\Delta R/R_o \approx \Delta G/G_o \propto \mathcal{V}_{mol}/\mathcal{V}_{pore}$ for $\mathcal{V}_{mol} \ll \mathcal{V}_{pore}$, where $G = R^{-1}$ and $\Delta G$ are the corresponding pore conductance and reduction. This simple relation has been extensively used for sizing particles from the amplitude of the induced transient resistive pulses.

This slicing technique (Eq. 1) is however not applicable to low aspect ratio pores ($d_p \gtrsim L_p$) for which resistive contributions from outside the pore cannot be ignored due to the significant electric field and voltage drop occurring in this access region. Although some models have addressed specific blockage scenarios,[16–21] there currently exists no generalizable approach akin to Equation 1 for modeling the conductance of partially blocked access regions that would be applicable to low aspect ratio channels, such as ultra-thin nanopores.[10] Such a model would be highly valuable in predicting electrical signatures from a wide range of translocating molecules, including cases of long obstructions that reside partially in the access regions even when inside the pore, or for obstructions entirely in the access regions, such as molecules approaching a pore.

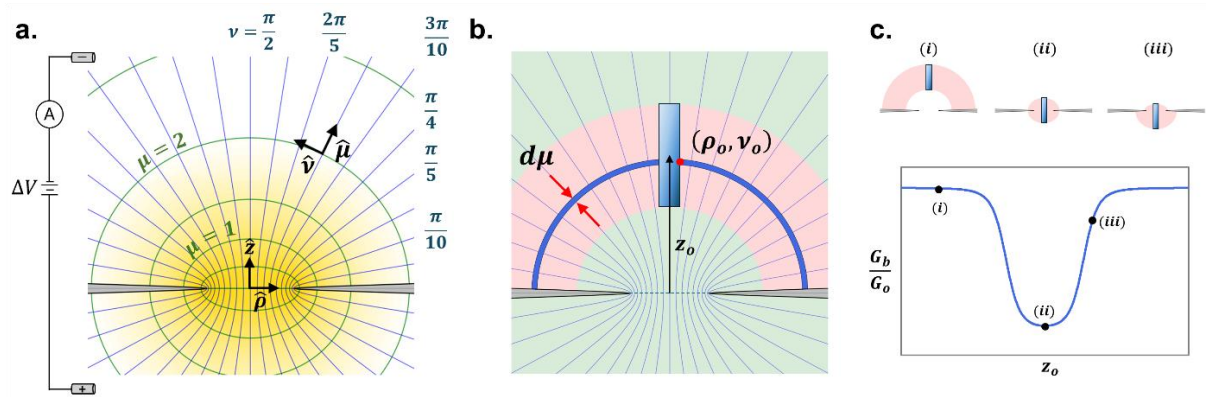

**Figure 1. a)** Oblate spheroidal coordinates used to map the access regions of a 2D pore. Constant-$\mu$ and constant-$\nu$ surfaces correspond to oblate spheroids and hyperboloid surfaces, respectively. Equipotential surfaces can be mapped to constant-$\mu$ oblate spheroids. **b)** Access region blocked by a cylindrical obstruction (at height $z_o$ above the pore) and divided into open (green) and blocked (red) oblate spheroidal segments. Blocked access resistance is obtained by partitioning the full region into spheroidal slices of thickness $d\mu$ (blue). **c)** Normalized conductance values during different translocation stages for a cylindrical obstruction through a 2D pore, with corresponding partitioning.

In our recent publication,[21] we introduced a highly accurate model for the conductance of access regions blocked by DNA-like cylindrical insulating obstructions, improving on the accuracy of the commonly used model of Kowalczyk *et al.*[17] which was shown to overestimate access conductance contributions by 100%. Here, we expand on this prior work and on existing modeling methods (e.g. Equation 1) to introduce a general framework for estimating the conductance of the access region of a channel in the presence of a wide range of insulating blocking objects (i.e. obstructions) by using thin oblate spheroidal slices to partition access regions (Figure 1b). We first demonstrate how this approach can be used to estimate the conductance of pores in 2D membranes in the presence of insulating obstructions with simple geometries such as cylinders, spheres, truncated cones, and spheroids at a distance $z_o$ from a pore (Figure 1c). These results can be useful for predicting electrical signatures from DNA, proteins and other translocating molecules with simple geometries. To address effects of off-axis passage, we derive numerically solvable expressions describing the radial dependence of cylinders and spheres located away from the pore's central axis. To illustrate the model's practical utility in modeling complex translocation blockades from molecules used in many sensing applications, we then show solutions for the resistance of more complex obstructions assembled from simple geometrical units. Finite-length pores of any aspect ratio can be modeled by using a combination of oblate spheroidal slices in the access regions, and circular slices inside the pore (Eq. 1), treating the access regions and the pore channel as resistors in series. Importantly, comparisons with finite element simulations and published experimental results are performed to assess the accuracy and limitations of the method, thus outlining conditions in which the model should be used. Due to the intricacy of most equations shown, a web-based tool is made available to predict the electrical signatures of any user-defined blocking object, which can be used in turn to optimize the nanopore architecture or molecular design that enhances the sensing resolution for a particular application.

**Oblate Spheroidal Framework for 2D membranes**

Following the work of others,[22–24] we use the oblate spheroidal variables $\mu \in [-\infty, \infty]$, $\nu \in [0, \pi/2]$ and $\phi \in [0, 2\pi]$ to map the access region of a pore, which are defined through the following transformations with cartesian coordinates $(\mu, \nu, \phi) \rightarrow (x, y, z)$:

$$\begin{aligned} x &= r_p \cosh\mu \cos\nu \cos\phi \\ y &= r_p \cosh\mu \cos\nu \sin\phi \\ z &= r_p \sinh\mu \sin\nu \end{aligned} \quad (2)$$

Note that the radial distance from the $z$ axis, $\rho$, can be calculated to be $\rho = r_p \cosh\mu \cos\nu$ (Fig. 1a). Importantly, by defining $r_p$ as the pore radius, this coordinate system naturally maps the access region: constant-$\mu$ surfaces form oblate spheroids with semi-axes of $r_p \sinh\mu$ along $\hat{z}$, and of $r_p \cosh\mu$ along $\hat{\rho}$, where the $\mu = 0$ surface corresponds to a disk of radius $r_p$ at the pore mouth, and the $\mu \rightarrow \pm\infty$ surfaces to infinitely large hemispheres.

To isolate and model the electrical response of access regions, we first study pores in 2D membranes ($L \rightarrow 0$) which by definition exclude contributions from the pore interior. For an open pore of radius $r_p$ in a 2D membrane, the Laplace equation $\nabla^2 V = 0$ can be solved exactly since the electric potential $V(\mu)$ depends solely on $\mu$ given that equipotential surfaces correspond to constant-$\mu$ oblate spheroids. Using boundary conditions of $V(\mu \rightarrow \pm\infty) = \pm\Delta V/2$ and $V(0) = 0$, corresponding to a potential difference applied across a circular electrode at the pore mouth and an infinitely large hemispherical electrode, an expression for the access region electric field is found:

$$\vec{E}_o(\mu, \nu) = -\frac{1}{h_\mu}\frac{dV}{d\mu}\hat{\mu} = \frac{\Delta V}{\pi r_p \cosh\mu \sqrt{\sinh^2\mu + \sin^2\nu}}\hat{\mu} \quad (3)$$

Here $h_\mu$ denotes one of the three scaling factors for oblate spheroidal coordinates, where $h_\mu = h_\nu = r_p\sqrt{\sinh^2\mu + \sin^2\nu}$ and $h_\phi = r_p \cosh\mu \cos\nu$. See Section S1 of the SI for a full derivation of Eq. 3.

Expanding on the method of Equation 1, the resistance of a pore in a 2D membrane can be calculated by first partitioning its access regions into an infinite number of oblate spheroidal slices of thickness $d\mu$.[25] As per Ohm's law, the infinitesimal resistance $dR(\mu)$ of each slice can be obtained by calculating the ratio of the potential difference across the slice, $dV(\mu) = \vec{E} \cdot d\vec{\ell}$, and the current through the slice, $I(\mu) = \int \int_S \sigma \vec{E} \cdot d\vec{S}$, where $\sigma$ denotes conductivity. A rigorous expression for the access resistance of a pore partially blocked by a translocating molecule thus requires an exact solution for $\vec{E}$, which in most scenarios is impossible to obtain. To estimate $dR(\mu)$, we introduce the assumption that $\vec{E} \approx \vec{E}_o = E_o \hat{\mu}$, i.e. the electric field is approximated by that of the open pore (Equation 3), as is commonly done for interior pore contributions.[10] If a current-blocking molecule's surface is delimited by $\nu = \nu_o(\mu, \phi)$ (Fig. 1b), then the infinitesimal resistance is approximated as:

$$dR(\mu) = \frac{\vec{E} \cdot d\vec{\ell}}{\int \int \sigma \vec{E} \cdot d\vec{S}} \approx \frac{E_o h_\mu d\mu}{\int_0^{2\pi} \int_0^{\nu_o(\mu,\phi)} \sigma E_o h_\nu h_\phi \, d\nu d\phi} = \frac{d\mu}{\int_0^{2\pi} \int_0^{\nu_o(\mu,\phi)} \sigma h_\phi \, d\nu d\phi} \quad (4)$$

Assuming a uniform conductivity and an insulating obstruction that is rotationally symmetric about the $z$-axis, the above expression can be integrated to estimate the resistance between two delimiting oblate spheroid surfaces:

$$R'_{obst}(\mu_1, \mu_2) = \int_{\mu_1}^{\mu_2} dR(\mu) = \frac{1}{2\pi\sigma r_p} \int_{\mu_1}^{\mu_2} \frac{\text{sech}\,\mu \, d\mu}{\sqrt{1 - \frac{\rho_o^2(\mu)}{r_p^2}\text{sech}^2\,\mu}} \quad (5)$$

Note in the above that $\rho_o(\mu)$ denotes the $\rho$-parametrization of the obstruction's surface, which delimits the region over which the current density $\sigma E$ is integrated (Fig. 1b). The expression for the resistance of an open region delimited by oblate spheroids intersecting $z = z_1$ and $z_2$ is readily obtained by setting $\rho_0 = 0$ in Eq. 5:

$$R_{free}(z_1, z_2) = \frac{1}{2\pi\sigma r_p}\left[\tan^{-1}\left(\frac{z_2}{r_p}\right) - \tan^{-1}\left(\frac{z_1}{r_p}\right)\right] \quad (6)$$

Importantly, note that in the open pore limit, we find $R_o(0, \infty) = 1/4\sigma r_p$ which corresponds exactly to the access resistance expression determined by Hall.[26]

As depicted in Figure 1b, by separating the access region into different blocked and open oblate spheroidal domains, the blocked-state conductance of a 2D pore in the presence of an obstruction $G_b$ of finite size can be calculated as the inverse of the sum of the resistances of the blocked interval, $R'_{obst}$, and the open segments above and below the insulating obstruction:

$$G_b = R_b^{-1} = \left[R_{free}^{bottom} + R'_{obst} + R_{free}^{top}\right]^{-1} \qquad (7)$$

and the corresponding conductance blockage is simply $\Delta G = G_o - G_b$, where $G_o = 2\sigma r_p$ is the open-pore conductance. Solving Equations 5-7 represents a general method by which to calculate the blocked-state conductance of pores in 2D membranes, which requires: i) determining the $\mu$ delimitations of blocked and open regions to use as integration bounds of Equations 5-6, ii) calculating the $\rho$-parametrization of the obstruction surface, $\rho_o(\mu)$, to find an expression for the integrand of $R'_{obst}$ (Equation 5), and iii) analytically or numerically calculating the solution to $R'_{obst}$, and inserting this result into the blocked-state equation (Equation 7). The rest of this work showcases how to evaluate Equations 5-7 for channels and translocating objects of various geometries.

**Cylindrical Obstruction**

To model the conductance of a 2D pore blocked by a rigid linear polymer like double-stranded DNA (dsDNA), we first consider an insulating cylinder of radius $r_c$ and of finite length $L_c$ and half-length $\ell_c \equiv L_c/2$ centered inside a 2D pore of radius $r_p$, as shown above Figure 2a. Here, the obstructed region is delimited by $\mu_\pm = \sinh^{-1}(\pm \ell_c/r_p)$, and the cylinder surface is trivially parametrized by $\rho_o(\mu) = r_c$. Inserting those values into Equation 5, the resistance of the segment obstructed by a finite-length cylinder $R'_{cyl}$ can be calculated:

$$R'_{cyl}(\ell_c, r_c) = \frac{1}{2\pi\sigma r_p} \int_{\sinh^{-1}\left(\frac{-\ell_c}{r_p}\right)}^{\sinh^{-1}\left(\frac{\ell_c}{r_p}\right)} \frac{\text{sech}\,\mu\, d\mu}{\sqrt{1 - \frac{r_c^2}{r_p^2}\text{sech}^2\mu}}$$

$$= \frac{1}{2\pi\sigma r_p} \int_{\frac{\pi}{2}-\tan^{-1}\left(\frac{\ell_c}{r_p}\right)}^{\frac{\pi}{2}-\tan^{-1}\left(\frac{-\ell_c}{r_p}\right)} \frac{d\theta}{\sqrt{1 - \frac{r_c^2}{r_p^2}\sin^2\theta}}$$

$$R'_{cyl}(\ell_c, r_c) = \frac{1}{2\pi\sigma r_p}\left[F\left(\frac{\pi}{2} - \tan^{-1}\left(\frac{-\ell_c}{r_p}\right), \frac{r_c}{r_p}\right) - F\left(\frac{\pi}{2} - \tan^{-1}\left(\frac{\ell_c}{r_p}\right), \frac{r_c}{r_p}\right)\right] \quad (8)$$

Note that Equation 8 was obtained using the variable substitution $\tanh\mu = \cos\theta$, and that the function $F(\varphi, k)$ denotes the incomplete elliptic integral of the first kind and is defined as $F(\varphi, k) = \int_0^\varphi (1 - k^2\sin^2\theta)^{-1/2}\, d\theta$. From Equation 8, a closed-form expression for the blocked state conductance of the 2D system, $G_b^{cyl}$, normalized by the corresponding open-pore conductance $G_o = 2\sigma r_p$, can be obtained by considering the additional contributions from the unobstructed oblate spheroidal segments (Eq. 6-7):

$$\frac{G_b^{cyl}}{G_o}(r_c, \ell_c) = \left[1 - \frac{2}{\pi}\tan^{-1}\left(\frac{\ell_c}{r_p}\right) + \frac{1}{\pi}F\left(\frac{\pi}{2} + \tan^{-1}\left(\frac{\ell_c}{r_p}\right), \frac{r_c}{r_p}\right) - \frac{1}{\pi}F\left(\frac{\pi}{2} - \tan^{-1}\left(\frac{\ell_c}{r_p}\right), \frac{r_c}{r_p}\right)\right]^{-1} \quad (9)$$

Figure 2a and its inset plot the dependence of $G_b^{cyl}/G_o$ on $r_c$ (for fixed length $L_c = 5r_p$), and of $G_b^{cyl}/G_o$ on $L_c$ (for fixed radius $r_c = 0.5r_p$), respectively. To evaluate the model accuracy, results from finite element simulations performed with identical system dimensions (see Methods) are plotted alongside predictions from Equation 9. Figure 2b plots the error percentage measured between simulations and Equation 9 and shows that the error monotonically increases as $r_c/r_p \to 1$, is highest for shorter $L_c$, but it is typically <5% for the practical case of dsDNA ($\geq$ 20 bp) through a $\geq$ 3 nm pore.

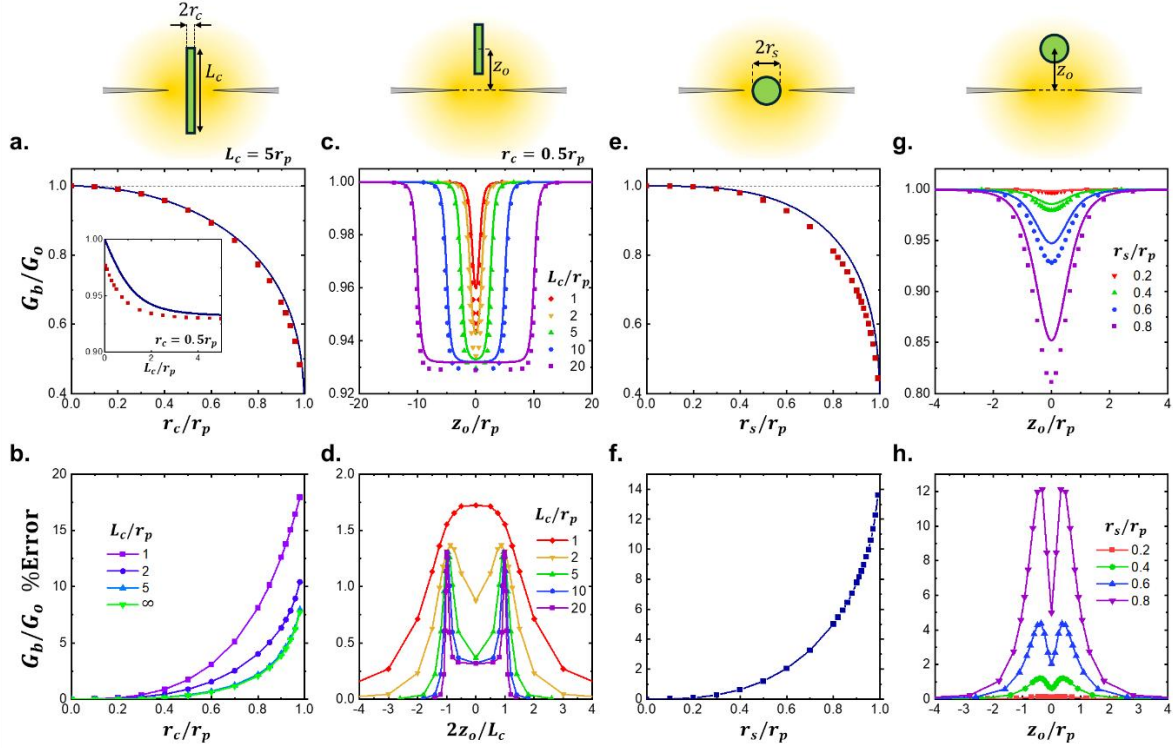

**Figure 2.** Dependence of normalized conductance $G_b/G_o$ on obstruction dimensions and distance from a 2D pore. Lines are from model predictions, whereas points are from finite element simulations. A normalized conductance of 1 is when $G_b = G_o$, i.e. the pore is unobstructed, whereas a value of 0 represents a fully blocked access region. **a)** $G_b/G_o$ vs. cylindrical obstruction radius $r_c$. Figure inset shows the dependence of $G_b/G_o$ on cylindrical obstruction length $L_c$. **b)** Absolute error percentage between modelled and simulated $G_b/G_o$ values from (a). **c)** $G_b/G_o$ vs. distance from pore $z_o$ for a cylindrical obstruction of radius $r_c = 0.5 r_p$ and different lengths $L_c$. **d)** Absolute error percentage between modelled and simulated $G_b/G_o$ values from (c). **e)** $G_b/G_o$ vs. centered ($z_o = 0$) spherical obstruction radius $r_s$. **f)** Absolute error percentage between modelled and simulated $G_b/G_o$ values from (e). **g)** $G_b/G_o$ vs $z_o$ for spherical obstructions of different radii. **h)** Absolute error percentage between modelled and simulated $G_b/G_o$ values from (g).

The normalized conductance blockage $\Delta G_{cyl}/G_o = (G_o - G_b^{cyl})/G_o$ can be calculated from Equation 9, from which it can be shown that $\Delta G_{cyl}/G_o$ scales proportionally to the ratio $r_c^2/r_p^2$ in the limit of ultra long and narrow cylindrical obstructions, i.e. $r_c \ll r_p \ll L_c$:

$$\frac{\Delta G_{cyl}}{G_o}(r_c \ll r_p \ll \ell_c) \approx \frac{1}{4}\frac{r_c^2}{r_p^2} = \frac{1}{3}\frac{2\pi r_c^2 r_p}{\frac{8}{3}\pi r_p^3} = \frac{1}{3}\frac{\mathcal{V}_{cyl}}{\mathcal{V}_{acc}} \qquad (10)$$

Equation 10 shows that just like for very long pores (Eq. 1), the fractional blockage is related to the ratio of cross-sectional areas $r_c^2/r_p^2$ and, equivalently, to the volume ratio $\mathcal{V}_{cyl}/\mathcal{V}_{acc}$. Here $\mathcal{V}_{acc}$ denotes the volume of the 2D-pore sensing region, defined as a constant-$\mu$ oblate spheroid with dimension $r_p$ along the $z$ axis, and $\mathcal{V}_{cyl}$ denotes the volume of the cylindrical obstruction contained within this sensing region (see Figure S1 of the SI).

Now, to model the conductance when cylindrically-shaped linear polymers like dsDNA are approaching a 2D pore of radius $r_p$, we consider an insulating cylinder of radius $r_c$ and of finite length $L_c$ along the pore's central $z$-axis, with its center located a distance $z_o$ above the pore, as depicted above Fig. 2c. Given the blocked region delimitation of $\mu_\pm = \sinh^{-1}\big((z_o \pm \ell_c)/r_p\big)$ and the trivial parametrization of $\rho_o(\mu) = r_c$, a closed-form expression for the conductance of the blocked system can be obtained with similar manipulations to those that led to Equation 9 (see Section S2 of the SI for the full derivation):

$$\frac{G_b^{cyl}}{G_o}(r_c, \ell_c, z_o) = \bigg[1 + \frac{1}{\pi}\tan^{-1}\bigg(\frac{z_o - \ell_c}{r_p}\bigg) - \frac{1}{\pi}\tan^{-1}\bigg(\frac{z_o + \ell_c}{r_p}\bigg)$$
$$+ \frac{1}{\pi}F\bigg(\frac{\pi}{2} + \tan^{-1}\bigg(\frac{z_o - \ell_c}{r_p}\bigg), \frac{r_c}{r_p}\bigg) - \frac{1}{\pi}F\bigg(\frac{\pi}{2} - \tan^{-1}\bigg(\frac{z_o + \ell_c}{r_p}\bigg), \frac{r_c}{r_p}\bigg)\bigg]^{-1} \quad (11)$$

Figure 2c plots the $z_o$-dependence of Equation 11, where the five plotted curves correspond to cylinder lengths of $L_c/r_p = 1, 2, 5, 10, 20$, and a fixed radius of $r_c = 0.5 r_p$. Data obtained from finite element simulations in identical conditions are color-matched and plotted as individual points alongside the corresponding model predictions (continuous lines). Figure 2d shows that Equation 11 is highly accurate, with consistent <2% error measurements.

We note that Equations 8-11 treat the cylinder extremities as oblate spheroidal surfaces instead of as flat disks. Although this treatment barely changes the results for the cases considered in Figure 2, Section S3 of the Supporting Information shows how to properly consider flat extremities, which additionally enables solutions for cylinders wider than the pore ($r_c > r_p$) within the access region, further generalizing the treatment of cylindrical obstructions.

**Spherical Obstruction**

To model the conductance of pores in the presence of insulating objects well approximated by spheres, such as viruses, nanoparticles, or globular proteins, we now consider a sphere of radius $r_s$ centered inside a 2D pore of radius $r_p$, as shown above Figure 2e. As detailed in Section S4 of the SI, by inserting the right boundaries and parametrization in Equations 5-7, a closed-form expression for the conductance of the 2D pore blocked by a spherical obstruction, $G_b^{sph}$, is obtained:

$$\frac{G_b^{sph}}{G_o}(r_s) = \left[1 - \frac{2}{\pi}\tan^{-1}\left(\frac{r_s}{r_p}\right) + \frac{2\,r_p}{\pi\,r_s}\tanh^{-1}\left(\frac{r_s^2}{r_p^2}\right)\right]^{-1} \tag{12}$$

Figure 2e shows the dependence of $G_b^{sph}/G_o$ on $r_s$ from Equation 12, and plots the corresponding values obtained from finite element simulations. The $r_s$ dependence of $G_b^{sph}/G_o$ is well captured by Equation 12, as shown in Figure 2f, which plots the error percentages between simulations and model predictions. Sub-15% error percentages were measured through the range of sphere radii explored ($r_s < 0.99 r_p$), notably with errors < 5% observed for spheres with $r_s < 0.8 r_p$.

As with long channels and cylindrical obstructions (Eq. 10), the normalized blockage $\Delta G_{sph}/G_o = (G_o - G_b^{sph})/G_o$ can be calculated from Eq. 12 and related to the volume ratio of the spherical obstruction and the access region volume $\mathcal{V}_{sph}/\mathcal{V}_{acc}$ for spheres with radii much smaller than that of the pore ($r_s \ll r_p$):

$$\frac{\Delta G_{sph}}{G_o}(r_s \ll r_p) \approx \frac{2}{3\pi}\frac{r_s^3}{r_p^3} = \frac{4}{3\pi}\frac{\frac{4\pi}{3}r_s^3}{\frac{8\pi}{3}r_p^3} = \frac{4}{3\pi}\frac{\mathcal{V}_{sph}}{\mathcal{V}_{acc}} \tag{13}$$

Similarly, it can be shown that the change in resistance introduced by a sphere passing through a 2D pore, $\Delta R_{sph}$, scales as $\Delta R_{sph} \propto r_s^3/r_p^4$, exactly as expected from long cylindrical channels.[9,10,12–14]

To model the access conductance as spherical objects approach and traverse 2D pores, we now consider an insulating sphere of radius $r_s$ whose center lies along the z-axis at a distance $z_o$ from a 2D pore of radius $r_p$ (Figure 2g). Section S4 details the derivation for the conductance of the obstructed 2D pore system, $G_b^{sph}(z_o, r_s)$:

$$\left(\frac{G_b^{sph}}{G_o}\right)^{-1} = 1 + \frac{1}{\pi}\tan^{-1}\left(\frac{z_o - r_s}{r_p}\right) - \frac{1}{\pi}\tan^{-1}\left(\frac{z_o + r_s}{r_p}\right)$$
$$+ \frac{1}{\pi}\int_{\sinh^{-1}\left(\frac{z_o-r_s}{r_p}\right)}^{\sinh^{-1}\left(\frac{z_o+r_s}{r_p}\right)} \frac{\text{sech}^2\mu \, d\mu}{\sqrt{1 - \frac{r_s^2 + z_o^2}{r_p^2}\text{sech}^2\mu + 2\frac{z_o^2}{r_p^2} - 2\frac{z_o}{r_p}\tanh\mu}\sqrt{1 + \frac{z_o^2}{r_p^2} - \frac{r_s^2}{r_p^2}\text{sech}^2\mu}} \quad (14)$$

Although Eq. 14 presents no closed form solution, it can be integrated numerically. Figure 2g shows the $z_o$ dependence of $G_b^{sph}/G_o$ for spheres with radii $r_s/r_p = 0.2, 0.4, 0.6, 0.8$ and plots finite-element simulations performed at various $z_o$ values. As in the $z_o = 0$ case, the error percentage calculated between simulations and model predictions is biggest for larger spheres, as shown in Figure 2h. Moreover, as with longer cylindrical obstructions (Fig. 2d), the error is not maximal for $z_o = 0$, but instead peaks at an intermediate value between $z_o = 0$ and $z_o = r_p$.

Note that Section S5 of the SI replots Figure 2 entirely but compares instead the blockage amplitudes $\Delta G/G_o$ predicted from the model with those calculated from simulations. Note also that the oblate spheroidal slicing method presented in this work is applicable to any rotationally symmetric obstruction whose surface is parametrizable by $v_o(\mu)$ or $\rho_o(\mu)$. To demonstrate the generality of the approach, the conductance of 2D pores in the presence of spheroidal and conical obstructions centered in the pore ($z_o = 0$) as well as positioned above the pore ($z_o \neq 0$) are derived and illustrated in Section S6 of the SI. Non-rotationally symmetric scenarios can also be considered, as demonstrated by Section S7 of the SI which shows the calculation for a wedged cylinder.

**Off-Axis Effects**

Partly due to the non-uniform electric field (Eq. 3) inside the pore, off-axis translocations are known to result in deeper blockades,[15,27–34] yet no mathematical treatment of this phenomenon currently considers access resistance contributions, which would be a beneficial tool for properly interpreting individual blockades and predicting blockage distributions of real-world experiments. Here, we utilize the oblate spheroidal slicing framework introduced above to model the off-axis effects of cylindrical and spherical obstructions positioned within a 2D pore.

To model the off-axis effects for 2D pores in the presence of rigid linear polymers such as dsDNA, we first consider an infinitely long cylindrical obstruction of radius $r_c$, whose central axis is located a distance of $\rho = r_o$ away from the pore center (Figure 3a). The equation for the $\phi$-dependent obstruction surface parametrization is shown in Section S8 of the SI, which demonstrates that the conditions $r_o \leq r_c$ and $r_o > r_c$ need to be treated separately, since the $\nu$ integration domain for $r_o \leq r_c$ is a single interval, whereas the $r_o > r_c$ scenarios result in two distinct intervals over which to integrate (Fig. 3a):

$$\left(\frac{G_b^{cyl}}{G_o}(r_o)\right)^{-1} = \begin{cases} 2\int_{-\infty}^{\infty} \frac{\text{sech}\,\mu}{\int_0^{2\pi}\sqrt{1-\frac{\rho_+^2}{r_p^2}(\phi)\,\text{sech}^2\,\mu\,d\phi}}\,d\mu, & \text{for } r_o \leq r_c \\ 2\int_{-\infty}^{\infty} \frac{\text{sech}\,\mu}{2\pi + \int_{-\sin^{-1}\left(\frac{r_c}{r_o}\right)}^{\sin^{-1}\left(\frac{r_c}{r_o}\right)}\left[\sqrt{1-\frac{\rho_+^2}{r_p^2}\text{sech}^2\,\mu} - \sqrt{1-\frac{\rho_-^2}{r_p^2}\text{sech}^2\,\mu}\right]d\phi}\,d\mu, & \text{for } r_o > r_c \end{cases} \quad (15)$$

Here $\rho_+$ and $\rho_-$ are the $\phi$-dependent surface parametrizations of the infinitely long cylinders (see Fig. 3a), the expressions for which are found in Equation S50 of the SI.

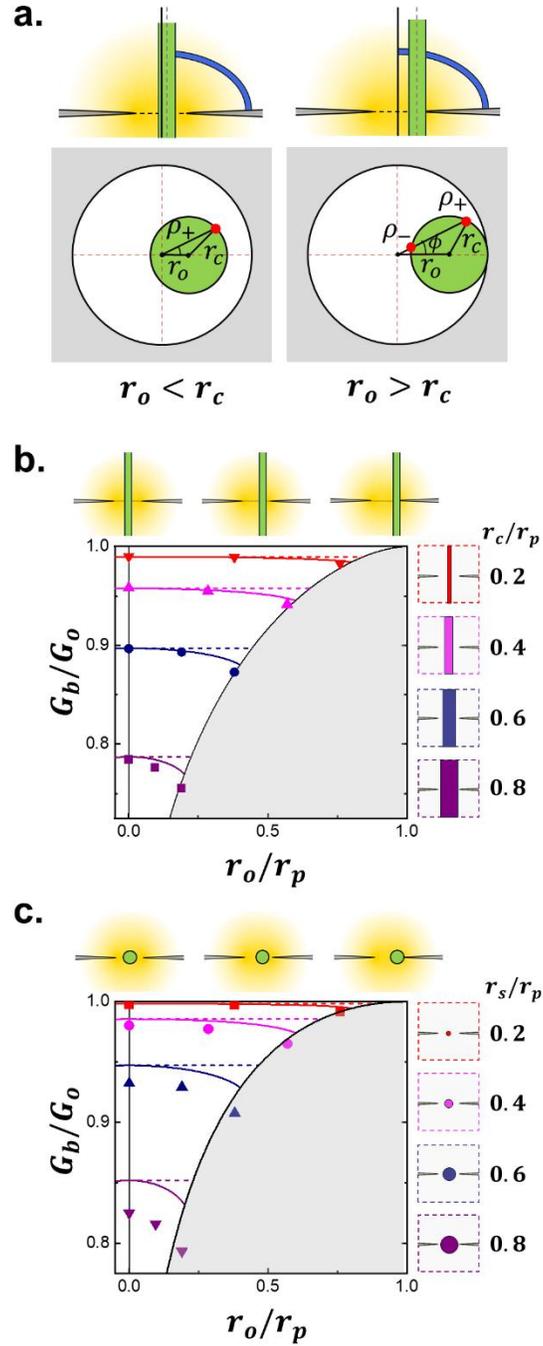

**Figure 3. a)** Schematic representation of 2D pores in the presence of off-centered obstructions at a distance $r_o$ from the pore center. **b)** Normalized conductance of 2D pore in blocked state versus normalized radial distance from the pore center of infinitely long cylindrical obstructions with different radii $r_c$ (20, 40, 60, 80% of pore radius $r_p$). The $r_o = 0$ case corresponds to a centered obstruction, while $r_o = r_p - r_c$ corresponds to the obstruction being in contact with the pore wall. **c)** Off-axis effects for vertically centered ($z_o = 0$) spherical obstructions of radius $r_s$. Continuous lines correspond to model predictions and individual points to calculations from finite element simulations. The greyed-out areas correspond to regions in the domain of Eqs. 15 where $r_o > r_p - r_{c,s}$.

Equation 15 can be integrated numerically, as displayed in Figure 3b which plots the dependence of the blocked state conductance $G_b^{cyl}/G_o$ on the radial position $r_o$ away from the pore center for different cylinder radii. Finite element simulations were performed to assess the accuracy of Equation 15 and are also plotted in Figure 3b. Equation 15 captures the off-axis dependence well, with larger $r_o$ values resulting in deeper blockades (lower $G_b$ values) for a given cylinder radius. For example, Eq. 15 predicts that a dsDNA fragment ($r_c = 1.1 \ nm$) inside a pore with $r_p = 5.5 \ nm$ would increase its blockage amplitude $\Delta G$ by ~50% going from the pore's center to the pore wall, with smaller pores experiencing smaller blockage amplitude fluctuations. A more quantitative comparison between simulations and Eq. 15 is shown in Section S8 of the SI.

Off-axis effects of spherical obstructions in 2D pores can also be modelled, though the manipulations are more involved due to the spherical surface parametrization $\rho_o(\mu, \phi)$ depending on both $\phi$ and $\mu$, unlike for cylinders. Figure 3c summarizes the results derived in Section S8 of the SI by plotting blockade state conductance values, $G_b^{sph}/G_o$, predicted by the model and simulated by finite element analysis for spheres of varying radii $r_s$, vertically centered ($z_o = 0$) and radially offset by $r_o$ within a pore of radius $r_p$. Although the absolute conductance values are less accurate for spheres than for cylinders, as in the $r_o = 0$ case (Figure 2), the increase in blockage amplitude $\Delta G$ with respect to $r_o = 0$ is better captured by the spherical obstruction model, as further discussed in Section S8 of the SI.

**Structured Obstruction**

Many pore sensing schemes have been developed that make use of DNA-protein structures or nanostructured DNA to create more uniquely identifiable signals or molecular barcodes.[35–40] Predicting the signals generated by the translocation of these complex molecular structures for a given set of pore dimensions is of high practical importance for successfully guiding experimental design. To this end, we now show how to address the conductance of complex obstructions approximated as a series of simpler obstruction units, as pictured in Figure 4a. This is achieved by further partitioning the obstructed oblate segment into multiple smaller segments, allowing the treatment of each obstruction sub-unit individually (Fig. 4a). For example, if $z_o$ is the vertical distance between the 2D pore of radius $r_p$ and the center of the complex molecule made up of $N$ simpler units and of total length $L_{comp} = 2\ell_{comp}$, then the conductance of the system is simply:

$$\frac{G_b^{comp}}{G_o}(z_o) = \left[1 + \frac{1}{\pi}\tan^{-1}\left(\frac{z_o - \ell_{comp}}{r_p}\right) - \frac{1}{\pi}\tan^{-1}\left(\frac{z_o + \ell_{comp}}{r_p}\right) + \sum_{i=1}^{N} R'_i(\mu_{i_-}, \mu_{i_+})\right]^{-1} \quad (16)$$

Here $R'_i(\mu_{i_-}, \mu_{i_+})$ denotes the resistance of the $i^{th}$ blocked segment delimited by $\mu_{i_-}$ and $\mu_{i_+}$, and other terms of Equation 16 originate from the open segment above and below the complex obstruction (Eq. 6). We note in particular that the treatment of complex obstructions made up of cylindrical sub-units of different sizes results in closed form solutions due to Eq. 9 being an analytic function.

Figures 4b-d show signals, i.e. $G_b^{comp}/G_o$ vs. $z_o$ traces, produced by the translocation of differently designed blocking structures as calculated from Eq. 16 (black) and compared to finite element simulations (red). Figure 4b shows the calculated electrical signature of twenty-three stacked cylinders designed to emulate the signal from a barcoded DNA design, i.e. eleven bit-like cylinders with radii $r_0 = 0.3r_p$ and $r_1 = 0.4r_p$ equally spaced along the backbone of a long cylinder of radius $r_c = 0.2r_p$. Similarly, Figure 4c shows the conductance signals of a sequence of eight spheres, with two of radius $r_s = 0.85r_p$ and six of $r_s = 0.7r_p$. To show that different sub-unit types can be used to make up complex obstructions, Figure 4d plots the conductance z-mapping of a series of cylinders ($r_c = 0.2r_p$, $L_c = 10r_p$) and spheres ($r_s = 0.5r_p$ and $0.7r_p$). As expected from the discussion surrounding isolated cylindrical and spherical obstructions (Figure 2), strong agreement is observed

between model predictions and simulations for smaller structured objects – predictions with larger structures deviate more significantly from the simulated conductance values (while still capturing well the shape of the expected traces).

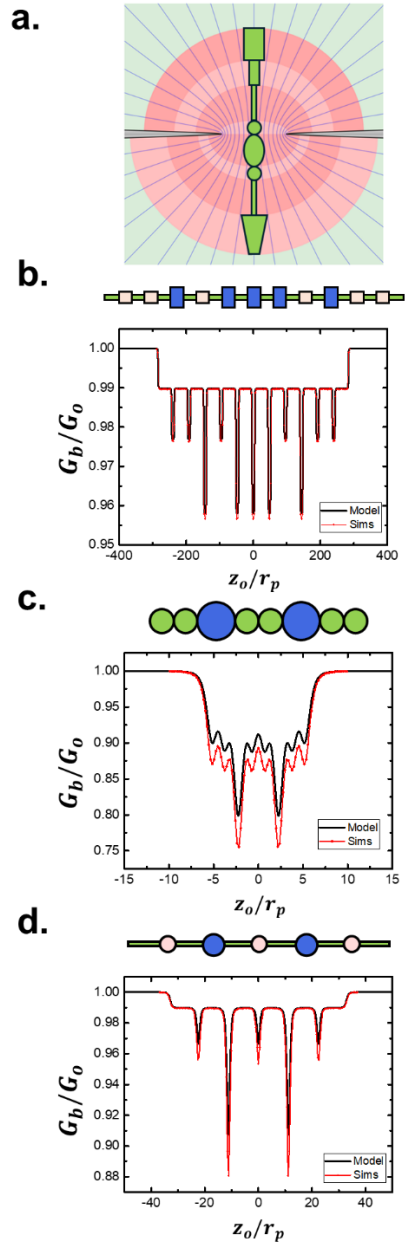

**Figure 4. a)** Oblate spheroidal partitioning of a structured obstruction made up of multiple simple geometrical units. **b)** z-dependence of $G_b/G_o$ for a sequence of cylinders, with three types of cylinders ($r_c, L_c = r_i, L_i$ for $i$ = 1,2,3 – see text for exact dimensions) mimicking the design of a DNA barcode. **c)** z-dependence of $G_b/G_o$ for a string of beads with $r_s = 0.85 r_p$ or $r_s = 0.7 r_p$ **d)** z-dependence of $G_b/G_o$ for a complex obstruction made up of cylinders ($r_c = 0.2 r_p, L_c = 10 r_p$) and spheres ($r_s = 0.5 r_p$ or $r_s = 0.7 r_p$).

To demonstrate the generality of Equation 16, Section S9 of the SI shows the traces of various complex blocking structures with different sequences of sub-unit geometries through 2D pores. After establishing the framework for modeling pores of finite length in the next section, we will further show that by modeling biomolecules as a sequence of simple sub-units in series, $G_b^{comp}$ vs $z_o$ traces can be obtained which, under a constant-velocity assumption, resemble temporal current traces measured experimentally.

**Finite-Length Nanopores**

In addition to modeling pore conductance in 2D membranes, oblate spheroidal slices can be used to treat the access resistances ($R_{acc}^{top}$ and $R_{acc}^{bot}$) of finite-length pores, as a natural extension of the circular slices used to calculate the resistance within a cylindrical channel ($R_{channel}$, see Eq. 1). This is of practical significance since many nanopore devices used in experimental work rely on relatively large ($r_p > 5\ nm$) nanopores fabricated in thin (0.3-20 nm) membranes, resulting in important access resistance contributions.[41–51] The general expression modeling the total resistance (access and channel) of an obstructed pore of finite length $L_p \equiv 2\ell_p$ and with a rotationally symmetric geometry described by a $z$-dependent radius $r_p(z)$ can thus be approximated by treating the two types of contribution in series:

$$R_b = R_{acc}^{bot} + R_{channel} + R_{acc}^{top}$$
$$= \frac{1}{2\pi\sigma r_p(-\ell_p)} \int_{-\infty}^{0} \frac{\mathrm{sech}\,\mu\, d\mu}{\sqrt{1 - \frac{\rho_o^2(\mu)}{r_p^2}\mathrm{sech}^2\mu}} + \frac{1}{\pi\sigma} \int_{-\ell_p}^{\ell_p} \frac{dz}{r_p^2(z) - r_o^2(z)}$$
$$+ \frac{1}{2\pi\sigma r_p(\ell_p)} \int_{0}^{\infty} \frac{\mathrm{sech}\,\mu\, d\mu}{\sqrt{1 - \frac{\rho_o^2(\mu)}{r_p^2}\mathrm{sech}^2\mu}} \tag{17}$$

Here, $\rho_o(\mu)$ and $r_o(z)$ denote the radial parametrization of the obstruction surface outside and inside the pore respectively (with the trivial case of $\rho_o(\mu) = r_o(z) = r_c$ for cylinders), and $\mu = 0$ corresponds to the membrane surface in the first and third integral term. Solving Equation 17 for a pore in the presence of an insulating molecule requires segmenting each access and channel region into blocked and open segments, naturally separating solutions of Equation 17 into a minimum of three regimes, depending on if the obstruction is: i) completely in one access region, ii) partially inside the pore and one of the access regions, or iii) completely inside the pore, if shorter than the pore, or partially in both access regions at once and fully inside the pore if longer than pore. Note that more conditions arise for non-symmetric pores (in $z$) or obstructions.

From Equation 17 and the corresponding treatment described above, closed-form $z_o$-mappings of cylindrical obstructions of length $L_c$ and radius $r_c$ passing through different rotationally symmetric channels of length $L_p$ (mimicking the time-varying translocation signatures of electrical measurements) can be obtained. For example, Figure 5a-c plots $z_o$-mappings of a cylindrical object ($r_c = 0.5r_p$ and $L_c = 20r_p$) passing through cylindrical, hyperboloidal, and conical shaped pores, the solutions of which are explicitly written out in Section S10 of the SI. Two mappings are shown per figure panel, corresponding to two different sets of dimensions tested for each pore shape. Corresponding conductance values calculated from finite element simulations are plotted as individual points alongside the $z_o$-mappings calculated from Equation 17. Generally, good agreement is found between model predictions and simulation values, although better agreement is found for pore radii that vary slowly along the $z$-axis, as is generally expected when working with circular slices (see Eq. 1).[9,25] Additionally, as with 2D membranes, numerical solutions can be obtained for spheres as well as any other obstruction geometry treated in previous sections, as calculated in Section S10 of the SI and as illustrated in Figure 5d, which shows the numerical solutions to Equation 17 for a sphere of radius $r_s = 0.5$ passing through cylindrical pores of two different lengths, with one being thinner than the sphere diameter (pink) and one longer (cyan).

Although only simple obstructions are shown in Figure 5, complex geometries can also be considered as per Equation 16 and Figure 4.

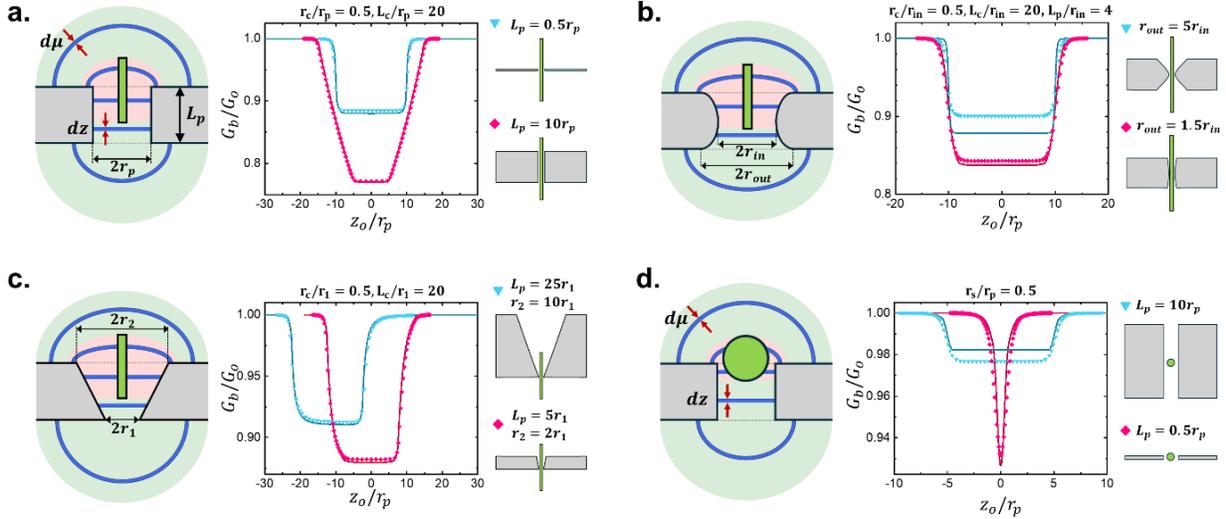

**Figure 5.** Conductance of finite length pores of various shapes during the translocation of a cylindrical $(r_c, L_c) = (0.5r_p, 20r_p)$ or spherical $(r_s = 0.5r_p)$ obstruction. Continuous lines correspond to model predictions (Eq. 17) and individual points to calculations from finite element simulations. **a)** Cylindrical obstructions through cylindrical pores with lengths $L_p = 0.5r_p$ (cyan inverted triangles) and $L_p = 10r_p$ (pink diamonds), showing different aspect ratios; **b)** Cylindrical obstructions through hyperboloidal pores with $L_p = 4r_{in}$ and external radii $r_{out} = 5r_{in}$ (cyan) and $r_{out} = 1.5r_{in}$ (pink), showing different pore tapers; **c)** Cylindrical obstructions through conical pores with dimensions $(L_p, r_2) = (5r_1, 2r_1)$ in pink, and $(L_p, r_2) = (25r_1, 10r_1)$ in cyan, emulating conical pores or pipette-like dimensions. **d)** A spherical obstruction going through cylindrical pores of length $L_p = 10r_p$ (cyan) and $L_p = 0.5r_p$ (pink), i.e. through membranes thicker and thinner than the sphere, respectively.

**Comparison with Experiments**

To assess the validity of the access resistance modeling method introduced in this work, we now compare model predictions to published experimental data acquired with nanopores made in quasi-2D material membranes, i.e., in ultra-thin, low aspect ratio pores, with diameters $d_p$ larger than their lengths $L_p$, where the applicability of the model is most relevant. Figure 6a reproduces data extracted from Garaj *et al.*[46] and plots the current blockages $\Delta I$ recorded for dsDNA translocations through pores of different diameters in graphene membranes ($L_p = 0.6\ nm$, as per the authors) in 3 M KCl solution ($\sigma = 27.5\ S/m$), under a bias of 160 mV. The model prediction for finite-length pores (Eq. 17) obtained by modeling pores as cylinders and DNA as an infinitely long cylinder with a 2.2 nm diameter is also plotted in Figure 6a, and reveals near-perfect agreement with the experimental blockage values. Similarly, Figure 6b reproduces data from Liu *et al.*[45] and, like the original work, plots the fractional blockage amplitude ($\Delta G/G_o$) versus the open-pore conductance $G_o$ measured for dsDNA translocations through pores formed in molybdenum disulfide membranes ($L_p = 1.4\ nm$, as per the authors) in 1 M KCl solution ($\sigma = 11\ S/m$). Model predictions (Eq.17) are plotted alongside the experimental data in red, again assuming cylindrical pore and obstruction geometries. Although Figure 6b shows less consistent agreement than for the graphene pores in 6a (perhaps due to additional electrokinetic effects present in lower salt concentrations), the reasonable agreements observed in Figures 6a-b show that the model presented herein captures the conductance blockage produced by the passage of dsDNA in ultra-thin pores well.

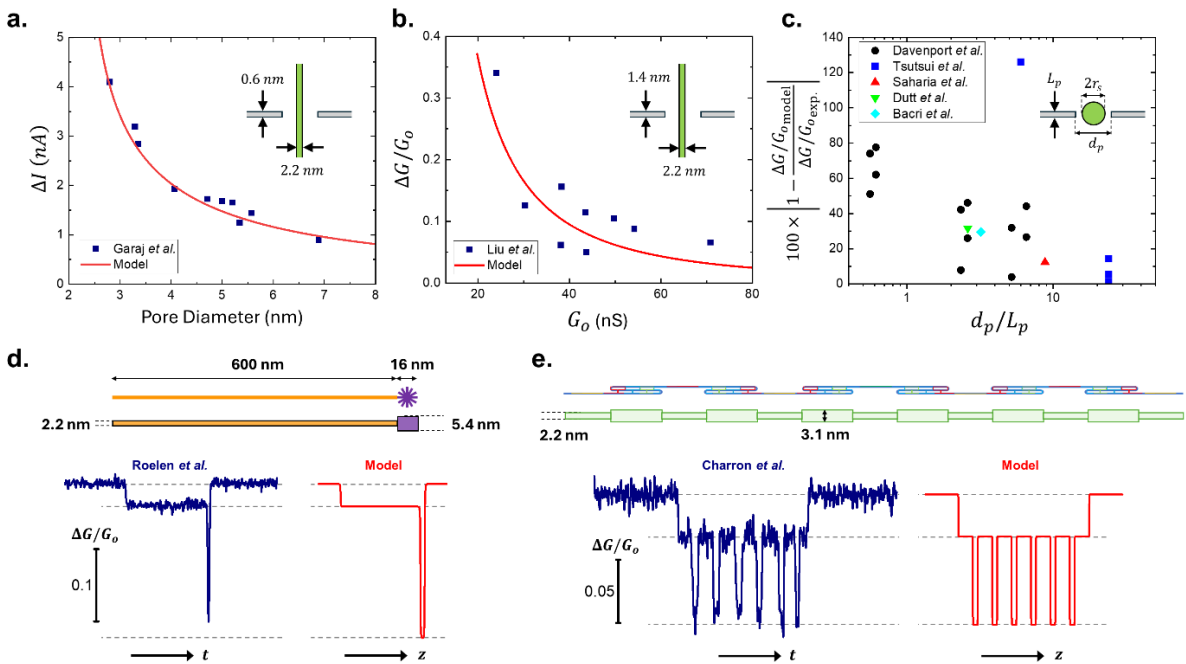

**Figure 6. Comparing model predictions to experimental data from low aspect-ratio nanopores. a)** Current blockage amplitude measured for dsDNA passing through pores of different sizes in graphene membranes from Garaj *et al.*[46] **b)** Fractional conductance blockage amplitude measured for dsDNA passing through pores of different sizes in molybdenum disulfide membranes from Liu *et al.*[45] **c)** Absolute percentage difference between published spherical nanoparticle- and protein-induced blockages[48–52] and corresponding model predictions measured in pores of different aspect ratios. **d)** Comparison of experimental temporal trace and model-predicted spatial trace for the current blockage of a shooting star DNA structure[53] approximated as two cylindrical obstructions in series for a cylindrical pore with $d_p = 8.9\ nm,\ L_p = 3.5\ nm$. **e)** Comparison of experimental temporal trace and model-predicted spatial trace for the current blockage of a DNA origami structure[54] approximated as two cylindrical obstructions of different sizes concatenated in series for a cylindrical pore with $d_p = 9.5\ nm,\ L_p = 6.8\ nm$.

To assess its ability to predict current blockages induced by the passage of spherical obstructions through finite length, yet still low-aspect nanopores, we now compare model predictions for cylindrical pores (Eq. 17) with published experimental current blockage data in such circumstances. Experimental blockage values for the translocations of silica and polystyrene nanoparticles were extracted from Davenport et al.,[48] Tsutsui et al.,[49] and Bacri et al.,[52] and blockages for the translocations of Human serum transferrin (hSTf) and Bovine Serum Albumin (BSA) proteins were extracted from Saharia et al.[51] and Dutt et al.[50] respectively. Model predictions were calculated by assuming cylindrical pore geometries, and setting the values of pore thickness and diameter, nanoparticle radius, as well as solution conductivity to those mentioned in the corresponding articles. Proteins were assumed to be spherical, with their cited hydrodynamic radius used for model calculations (See Section S11 of the SI for more details on experimental conditions and values used for the calculations). Figure 6c displays the absolute percentage difference between fractional blockages from experimental measurements, $\Delta G/G_{o_{exp}}$, and model predictions, $\Delta G/G_{o_{model}}$, plotted against the pore's aspect ratio, i.e. the ratio of pore diameter to pore length $L_p/d_p$. Despite the low salt concentrations used in the nanoparticle experiments (0.1 M KCl,[48] 0.4x PBS,[49] 0.01 M KCl[52]) and the limitations of our model that ignores electrokinetic effects, a surprisingly decent agreement (≲ 40% error) is found for higher aspect ratio pores, i.e. higher values of $d_p/L_p$. The more significant disagreement for lower $d_p/L_p$ values likely arises due to the geometry of thicker pores differing from the cylindrical geometry assumed in the model. Section S11 of the SI provides further details on the experimental data and model parameters shown in Figure 6 and presents additional nanoparticle blockage data[44] which differ more significantly (>1000%) from our model predictions, likely due to the signals being attenuated by the limited temporal resolution of the experimental system.

Lastly, Figures 6d-e highlight how the blockage signals of DNA nanostructures translocating though finite length pores can be predicted accurately by approximating their shapes as a sequence of simple geometric objects, as per Figure 4, and combining the circular and oblate spheroidal slicing methods for the pore interior and exterior, respectively (Eq. 17). Figure 6d shows the current vs. time trace (blue) from the passage of a shooting star DNA structure[53] through a pore with effective diameter and length of 8.9 nm and 3.5 nm respectively, as calibrated from conductance data employing a dsDNA fragment as a molecular ruler.[21] The shooting star structure is made up of a 1.8 kbp dsDNA tail attached to a 12-arm (24 bp each) dsDNA star. To predict the shape of its current trace, as shown in red in Figure 6d, the nanostructure's geometry was approximated as two connected cylinders. A cylinder of diameter 2.2 nm and length $1800\ bp \times 0.34\ nm/bp = 612\ nm$ was used to model the dsDNA tail, and a second cylinder of diameter $\sqrt{6} \times 2.2\ nm \approx 5.4\ nm$ and length $48\ bp \times 0.34\ nm/bp = 16.3\ nm$ to model the conformation of the star segment with six pairs of its dsDNA strands aligned in opposite directions, as reported in earlier work.[55] Similarly, Figure 6e shows the experimental temporal trace recorded (blue) and the predicted spatial trace calculated (red) for the passage of a structured DNA origami structure[54] into a nanopore with effective calibrated diameter of 9.5 nm and thickness of 6.8 nm.[21] The DNA nanostructure is made from a single-stranded DNA scaffold locally folded on itself at six locations, held by ~171 short staple oligomer strands, as described in more detail elsewhere.[54] The DNA structure is therefore made up of interspaced domains that can be approximated as a series of cylinders with diameters of 2.2 nm (dsDNA segments) or $\sqrt{3} \times 2.2$ nm (folded domains) in order to predict the shape of the current trace, as shown in Figure 6e.

Figures 6d and 6e show that the shape of the temporal current trace of a translocation event through finite length pores can be well predicted using the improved conductance model introduced in this work. The model can also provide simple physical intuitions about the approximated geometry of structures during nanopore translocation. Both of these features can be highly valuable for designing and interpreting nanopore experiments. For this reason, we have made a web-based tool implementing Equations 16 and 17 to predict the conductance signals of any user-defined blocking structure in pores of 2D, cylindrical, conical or hyperbolic geometries.[56] The online tool also includes the ability to add noise and filter effects to produce more realistic signals. Such tools, we believe, can assist in designing structured biomolecules to be efficiently sensed with nanopores, or conversely, in optimizing pore diameter and shape to better resolve a given molecular design.

**Applicability of Modeling Framework**

With the accuracy and flexibility of the access resistance modeling framework demonstrated across various scenarios, we now highlight its underlying assumptions and error sources to better inform on its applicability and limitations. Calculating the resistance of a conducting volume by integrating the resistances of volume-partitioning slices is a well-established method dating back to Maxwell's treatise.[25] The method is known to be exact if the slice surfaces correspond to equipotential surfaces, and to underestimate the resistance otherwise. The oblate spheroidal slices used in the modeling technique correspond exactly to the equipotential surfaces of a 2D pore in its open state, but differ in its blocked state, explaining why conductance blockage predictions are consistently shallower than finite element analysis calculations (Figures 1-4). We note however that shape correction factors can be developed to increase model accuracy by accounting for the deflection of field lines around insulating obstructions, which are otherwise ignored when integrating over open-pore equipotential surfaces.[11,14,15,33] For finite-length pores treated in Figure 5 and the web-app,[56] circular disks are used to model the pore interior, as is commonly done which is known to be more accurate for pore geometries resembling cylinders, and less accurate for steeply curved pore surfaces found in sharp conical and hourglass shaped pores for example. Moreover, in the transition region between the inside and outside of the pore, the equipotential surfaces deviate from being purely circular and oblate spheroidal, respectively, resulting in both open and blocked states being non-exact and underestimated. In addition to these inexact field treatments, we note that the model ignores electrokinetic phenomena that can arise, especially in experiments performed in low salt conditions (< 1 M concentrations), partly due to charged pore surfaces or charged translocating molecules.[57,58] Lastly, many obstructions present either intricate closed-form solutions or solutions that require numerical integration. This prompted the creation of the web-based tool aiming to facilitate potentially arduous computations. Although more advanced computational approaches can model ionic current responses with atomistic precision while considering electrokinetic effects,[59] this tool is useful for designing optimal experimental conditions or molecular constructs by coarsely and rapidly predicting current signatures while maintaining good accuracy for many practical experimental conditions. Additionally, unlike simulation approaches, the introduced modeling framework gives physical insight into the system studied, as per the scaling of conductance blockage with obstruction and pore dimensions (Equations 10 and 13), in addition to predicting conductance blockage values.

In summary, through the oblate spheroidal partitioning of the space outside a pore, we have introduced a general framework by which the resistance of the access regions blocked by an object can be modeled. This approach presents itself as a natural extension to the circular slice method commonly used to model resistances of channel interiors (Eq. 1).[10] The applicability of the method is exemplified by the various scenarios (obstruction and pore geometries) and physical phenomena (e.g. off-axis effects) it successfully estimates, as assessed through comparison to finite element analyses and experimental translocation data measured using low aspect ratio nanopores. Of particular interest, the modeling of simple or complex molecular structures as a sequence of geometrically simple blocking units (i.e., cylinders, spheres, etc.), allows the calculation of conductance traces as a function of the distance from the pore, useful for predicting the shape of the current traces from actual experiments. The methods presented herein represent a very useful first step towards obtaining more accurate and generally applicable models for blocked access regions, recently described as being in their infancy.[10] Future steps will include determining the obstruction-specific shape correction factors required to transform the approximations from this work into exact expressions, as previously achieved for long channels, allowing optimal accuracy for blockage prediction and molecular sizing.[11,14,15,33] With rapidly advancing micro- and nano-fabrication techniques resulting in thinner membranes with access-resistance dominated sensing conditions, we believe the introduced technique and associated web-based tool can greatly assist researchers in sizing molecules, for protein fingerprinting and nanoparticle sensing applications, or for optimally designing complex biomolecule barcodes in multiplexing applications.

**Methods**

*Simulations:* Finite element simulations were performed with the Electric Currents Module of COMSOL Multiphysics 5.1. In its open state, a nanopore was modeled as a channel of radius $r_p$ in an insulating membrane of either zero or finite thickness $L_p$ (a.k.a. pore length), centred inside a cylindrical meshing domain of length 10 µm and diameter 5 µm. In its blocked state, an insulating region with dimensions matching those of the particular obstruction being modeled was added to the system at the desired coordinates ($r_o$, $z_o$). Solutions to Poisson's equation were found within the finite-resistivity regions (defined by constant conductivity $\sigma$) of a given geometry, using boundary conditions of 0.2 and 0 V at the top and bottom surfaces of the meshing domain, respectively ($\Delta V =$

$V_{top} - V_{bottom}$). The resulting electrical conductance of the system was evaluated in both open and blocked states such that the numerical solver converged as the mesh size was reduced. The normalized conductance $G/\sigma$ was calculated by integrating the current density $J$ across the planar top boundary to find the total current $I$ through the system and then dividing this value by the defined transmembrane potential ($G = I/\Delta V$) and conductivity.

*Experimental Data Extraction:* The experimental data contained in Figure 6 were either taken from the text in their respective publications, or estimated visually from the original figures using pixel counts.

**Acknowledgments**

M. C. acknowledges support of the Ontario Graduate Scholarship (OGS). All authors would like to acknowledge the support of the Natural Sciences and Engineering Research Council of Canada (NSERC), [funding reference number RGPIN-2021-04304], and of Oxford Nanopore Technologies.

**Data Availability**

The authors declare that the data supporting the findings of this study are available within the paper, its supplementary information files. To reproduce the results readers can use the freely available web-app (https://www.tcossalab.net/signalGenerator).

# Supplementary Information for: Predicting Resistive Pulse Signatures in Nanopores by Accurately Modeling Access Regions

Martin Charron, Zachary Roelen, Deekshant Wadhwa, Vincent Tabard-Cossa*

150 Louis-Pasteur Private, Department of Physics, University of Ottawa, Ottawa K1N 6N5, Canada

*Corresponding Author: tcossa@uottawa.ca

**S1. Electric Field in a 2D pore**

**S2. Finite Cylindrical Obstruction – z dependence**

**S3. Treatment of Flat Cylindrical Obstruction Extremities**

**S4. Spherical Obstruction**

**S5. Fractional Blockage Error**

**S6. Rotationally Symmetric Obstructions – More Examples**

**S7. Rotationally Asymmetric Obstructions – Wedged Cylinder**

**S8. Rotationally Asymmetric Obstructions – Off-Axis Effects**

**S9. Complex Obstruction – Modeling Molecules**

**S10. Finite-Length Pore Equations**

**S11. Experimental Comparison Details**

## S1. Electric Field in a 2D pore

Oblate spheroidal coordinates $(\mu, \nu, \phi)$ are described by the following transformations with cartesian coordinates $(x, y, z)$:

$$\begin{aligned} x &= r_p \cosh\mu \cos\nu \cos\phi \\ y &= r_p \cosh\mu \cos\nu \sin\phi \\ z &= r_p \sinh\mu \sin\nu \end{aligned} \tag{S1}$$

As discussed in the main text, the electric potential $V$ between a disk electrode of radius $r_p$ and an infinitely large hemispherical electrode is assumed to depend solely on $\mu$. Under the boundary conditions $V(0) = 0$ and $V(\pm\infty) = \pm\Delta V/2$, and assuming no net charge in the system, a solution for $V(\mu)$ can be found by solving the curvilinear Laplacian equation:

$$\begin{aligned} 0 &= \nabla^2 V \\ 0 &= \frac{1}{h_\mu h_\nu h_\phi} \frac{\partial}{\partial \mu}\left(\frac{h_\nu h_\phi}{h_\mu} \frac{\partial V}{\partial \mu}\right) \\ 0 &= \frac{1}{r_p^3 (\sinh^2\mu + \sin^2\nu) \cosh\mu \cos\nu} \frac{\partial}{\partial\mu}\left(r_p \cosh\mu \cos\nu \frac{\partial V}{\partial\mu}\right) \\ 0 &= \frac{\partial}{\partial\mu}\left(\cosh\mu \frac{\partial V}{\partial\mu}\right) \\ c_1 &= \cosh\mu \frac{\partial V(\mu)}{\partial\mu} \\ V(\mu) &= c_1 \int \frac{d\mu}{\cosh\mu} \\ V(\mu) &= c_1 \tan^{-1}(\sinh\mu) + c_2 \end{aligned}$$

$$V(\mu) = \frac{\Delta V}{\pi} \tan^{-1}(\sinh\mu) \tag{S2}$$

Note here that $h_\mu, h_\nu$ and $h_\phi$ are scaling factors defined as $h_u = \sqrt{\left(\frac{\partial x}{\partial u}\right)^2 + \left(\frac{\partial y}{\partial u}\right)^2 + \left(\frac{\partial z}{\partial u}\right)^2}$ for $u = \mu, \nu, \phi$, evaluated as:

$$\begin{aligned} h_\mu &= r_p \sqrt{\sinh^2\mu + \sin^2\nu} \\ h_\nu &= r_p \sqrt{\sinh^2\mu + \sin^2\nu} \\ h_\phi &= r_p \cosh\mu \cos\nu \end{aligned} \tag{S3}$$

The electric field can be calculated from Equations S2 and S3:

$$\vec{E}(\mu, \nu) = -\frac{1}{h_\mu}\frac{\partial V}{\partial \mu}\hat{\mu} = -\frac{\Delta V}{\pi r_p}\frac{\text{sech}\,\mu}{\sqrt{\sinh^2\mu + \sin^2\nu}}\hat{\mu} \tag{S4}$$

Note that since $V(\mu)$ depends only on $\mu$, the electric field is directed along $\hat{\mu}$ (See Figure 1a of main text), and its magnitude $E_o = \|\vec{E}\|$ depends on both $\mu$ and $\nu$, due to the introduction of the $h_\mu$ scaling factor in Equation S3.

## S2. Finite Cylindrical Obstruction – z dependence

Consider an insulating cylinder of radius $r_c$ and of finite length $L_c = 2\ell_c$ with its center located a distance $z_o$ away from a 2D pore of radius $r_p$. Inserting the obstructed region delimitation of $\mu_\pm = \sinh^{-1}\left((z_o \pm \ell_c)/r_p\right)$ and the trivial parametrization of $\rho_o(\mu) = r_c$ into the expression for the obstructed segment (Equation 5 in main text) results in:

$$R'_{cyl}(r_c, \ell_c, z_o) = \frac{1}{2\pi\sigma r_p} \int_{\sinh^{-1}\left(\frac{z_o-\ell_c}{r_p}\right)}^{\sinh^{-1}\left(\frac{z_o+\ell_c}{r_p}\right)} \frac{\mathrm{sech}\,\mu}{\sqrt{1 - \frac{r_c^2}{r_p^2}\mathrm{sech}^2\,\mu}}\, d\mu$$

$$= \frac{1}{2\pi\sigma r_p} \int_{\frac{\pi}{2}-\tan^{-1}\left(\frac{z_o+\ell_c}{r_p}\right)}^{\frac{\pi}{2}-\tan^{-1}\left(\frac{z_o-\ell_c}{r_p}\right)} \frac{d\theta}{\sqrt{1 - \frac{r_c^2}{r_p^2}\sin^2\theta}}$$

$$= \frac{1}{2\pi\sigma r_p}\left[F\left(\frac{\pi}{2} - \tan^{-1}\left(\frac{z_o - \ell_c}{r_p}\right), \frac{r_c}{r_p}\right) - F\left(\frac{\pi}{2} - \tan^{-1}\left(\frac{z_o + \ell_c}{r_p}\right), \frac{r_c}{r_p}\right)\right] \quad (S5)$$

Equation S5 is obtained from the variable substitution $\tanh\mu = \cos\theta$, resulting in $\mathrm{sech}\,\mu\, d\mu = -d\theta$, and from identities of inverse trigonometric functions:

$$\begin{aligned}
\theta|_{\mu=\sinh^{-1}(x)} &= \cos^{-1}(\tanh(\sinh^{-1}(x))) \\
&= \cos^{-1}\left(\frac{x}{\sqrt{1+x^2}}\right) \\
&= \frac{\pi}{2} - \sin^{-1}\left(\frac{x}{\sqrt{1+x^2}}\right) \\
&= \frac{\pi}{2} - \tan^{-1} x
\end{aligned} \quad (S6)$$

As noted in the main text the function $F(\varphi, k)$ is the incomplete elliptic integral of the first kind and is defined as $F(\varphi, k) = \int_0^\varphi (1 - k^2 \sin^2\theta)^{-1/2}\, d\theta$.

The resistance of the unobstructed segments above and under the cylinder can be calculated:

$$\begin{aligned}
R_{free}(z_o - \ell_c, z_o + \ell_c) &= R_{free}^{bot} + R_{free}^{top} \\
&= \frac{1}{2\pi\sigma r_p}\int_{-\infty}^{\sinh^{-1}\left(\frac{z_o-\ell_c}{r_p}\right)} \mathrm{sech}\,\mu\, d\mu + \frac{1}{2\pi\sigma r_p}\int_{\sinh^{-1}\left(\frac{z_o+\ell_c}{r_p}\right)}^{\infty} \mathrm{sech}\,\mu\, d\mu \\
&= \frac{1}{2\pi\sigma r_p}[\tan^{-1}(\sinh\mu)]_{-\infty}^{\sinh^{-1}\left(\frac{z_o-\ell_c}{r_p}\right)} + \frac{1}{2\pi\sigma r_p}[\tan^{-1}(\sinh\mu)]_{\sinh^{-1}\left(\frac{z_o+\ell_c}{r_p}\right)}^{\infty} \\
&= \frac{1}{2\pi\sigma r_p}\left(\tan^{-1}\left(\frac{z_o - \ell_c}{r_p}\right) + \frac{\pi}{2}\right) + \frac{1}{2\pi\sigma r_p}\left(\frac{\pi}{2} - \tan^{-1}\left(\frac{z_o + \ell_c}{r_p}\right)\right) \\
&= \frac{1}{2\sigma r_p}\left[1 + \frac{1}{\pi}\tan^{-1}\left(\frac{z_o - \ell_c}{r_p}\right) - \frac{1}{\pi}\tan^{-1}\left(\frac{z_o + \ell_c}{r_p}\right)\right] \quad (S7)
\end{aligned}$$

From Equations S4 and S6, an expression for the conductance of the entire 2D pore system in the presence of a cylindrical obstruction of finite length is found:

$$\frac{G_b^{cyl}}{G_o}(r_c, \ell_c, z_o) = \left[1 + \frac{1}{\pi}\tan^{-1}\left(\frac{z_o - \ell_c}{r_p}\right) - \frac{1}{\pi}\tan^{-1}\left(\frac{z_o + \ell_c}{r_p}\right)\right.$$
$$\left. + \frac{1}{\pi}F\left(\frac{\pi}{2} - \tan^{-1}\left(\frac{z_o - \ell_c}{r_p}\right), \frac{r_c}{r_p}\right) - \frac{1}{\pi}F\left(\frac{\pi}{2} - \tan^{-1}\left(\frac{z_o + \ell_c}{r_p}\right), \frac{r_c}{r_p}\right)\right]^{-1} \quad (S8)$$

Here $G_o = 2\sigma r_p$ is the open 2D pore conductance. Figure S2 plots Equation S8 against the normalized distance $z_o/(\ell_c + r_p)$, for which a value of $\pm 1$ corresponds the closest extremity of the cylinder being at a distance $r_p$ from the pore mouth. Although each trace corresponds to a different cylinder length, all traces appear to converge near $z_o = \ell_c + r_p$, thus indicating that the dimension of the sensing volume of a 2D pore is on the order of $r_p$ along the z-axis.

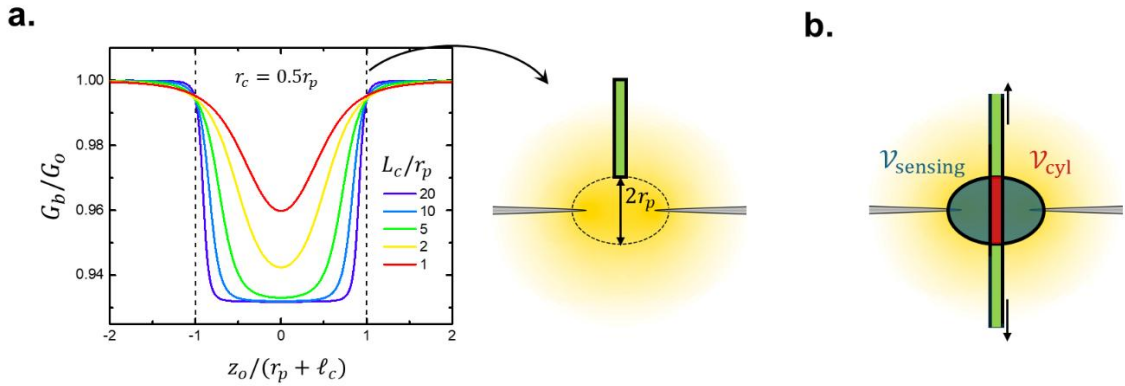

**Figure S1. a)** z-dependence of 2D pore conductance in the presence cylindrical obstructions of various lengths $L_c$, of fixed radius $r_c = 0.5r_p$, with their center at a distance $z_o$ from the pore center. **b)** Sketch demonstrating the interpretations of $\mathcal{V}_{\text{sensing}}$ and $\mathcal{V}_{\text{cyl}}$ as the volumes of the 2D pore sensing volume, and the volume of the cylinder contained with it, respectively.

Additionally, when evaluated for long cylinders ($\ell_c \gg r_p$) centered inside a pore ($z = 0$), Equation S8 reduces to a simpler form involving the complete elliptic integral of the first kind $K(k)$, related to the incomplete elliptic integral of the first kind through $K(k) = F\left(\frac{\pi}{2}, k\right)$.

$$\frac{G_b^{cyl}}{G_o}(r_c, \ell_c \gg r_p, z_o = 0) = \left[\frac{1}{\pi}F\left(\pi, \frac{r_c}{r_p}\right)\right]^{-1}$$
$$= \left[\frac{2}{\pi}K\left(\frac{r_c}{r_p}\right)\right]^{-1}$$
$$= \left[1 + \frac{1}{4}\frac{r_c^2}{r_p^2} + \frac{9}{64}\frac{r_c^4}{r_p^4} + \cdots\right]^{-1}$$
$$= 1 - \frac{1}{4}\frac{r_c^2}{r_p^2} - \frac{9}{64}\frac{r_c^4}{r_p^4} + \cdots \quad (S9)$$

The final expression of Equation S9 used the known relationship $2K(k) = F(\pi, k)$ and the series expansion of $K(k)$ around $k = 0$, the first terms of which are written.

For small values of cylinder radii, $r_c \ll r_p$, Equation S9 indicates that the fractional conductance blockage $\Delta G/G_o = 1 - G_{cyl}/G_o$ is proportional to $r_c^2/r_p^2$:

$$\frac{\Delta G_{cyl}}{G_o}(r_c \ll r_p \ll \ell_c, z_o = 0) \approx \frac{1}{4}\frac{r_c^2}{r_p^2} = \frac{1}{3}\frac{2\pi r_c^2 r_p}{\frac{8}{3}\pi r_p^3} = \frac{1}{3}\frac{\mathcal{V}_{cyl}}{\mathcal{V}_{sensing}} \tag{S10}$$

Multiplying $r_c^2$ and $r_p^2$ by $2\pi r_p/3$, we note interestingly that the fractional blockade $\Delta G/G_o$ is proportional to the ratio of $\mathcal{V}_{cyl}$ and $\mathcal{V}_{sensing}$, where $\mathcal{V}_{sensing}$ is the volume of the oblate spheroid with semi-axis $r_p$ along $z$ (which from now on we designate as the "sensing region"), and $\mathcal{V}_{cyl}$ is the partial volume of the cylinder contained within the sensing volume.

**S3. Treatment of Flat Cylindrical Obstruction Extremities**

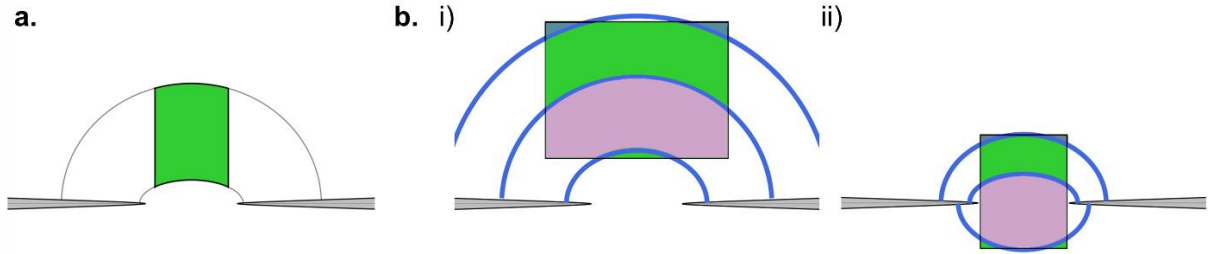

**Figure S2**. **a)** Sketch showing how cylindrical obstructions are treated as having oblate spheroidal extremities in the main text. **b)** Sketches demonstrating the two independent treatments required to address flat extremities of cylindrical obstructions outside the pore (i) or inside the pore (ii).

The treatment of cylindrical obstructions in the main text inherently assumes cylinder extremities are curved and follow the shape of the oblate spheroidal surface delimiting the obstructed segment, as depicted in Figure S2a. Cylinders with radii similar in size to that of the pore therefore have their extremities significantly deformed, with the scenario of $r_c > r_p$ even being untreatable. A more thorough treatment of cylindrical obstructions requires segmenting the access region into more domains, depending on whether constant-$\mu$ oblate spheroidal slices intersect with the cylinder's bottom, side, top, or any combination of those, as demonstrated in Figure S2b. Two different scenarios must be considered: i) the cylinder is completely outside the pore or ii) the cylinder is partially inside the pore.

<u>Outside the pore ($|z_o| > \ell_c$)</u>

When the cylinder is completely outside the pore, as displayed in Figure S2b-i, the obstructed volume has to be separated into three independently treated oblate spheroidal segments corresponding to resistances $R'^{bot}_{obs}$, $R'^{side}_{obs}$, and $R'^{side-top}_{obs}$, in addition to the unobstructed resistive contributions:

$$G_b^{cyl}(r_c, \ell_c, z_o) = \left[R_{free}^{bot} + R'^{bot}_{cyl} + R'^{side}_{cyl} + R'^{side-top}_{cyl} + R_{free}^{top}\right]^{-1} \tag{S11}$$

Before determining expressions for each term in Eq. S10, let us first determine the values of $\mu$ delimiting the different integration domains. Generally, for a cylinder edge located at $z$, this is achieved by calculating for what value of $\mu$ a constant-$\mu$ oblate spheroid intersects with a cylinder edge $(\rho, z) = (r_c, z)$:

$$\frac{r_c^2}{\cosh^2 \mu} + \frac{z^2}{\sinh^2 \mu} = r_p^2 \tag{S12}$$

Equation S12 can be rewritten as

$$0 = r_p^2 (\sinh^2 \mu)^2 + (r_p^2 - r_c^2 - z^2) \sinh^2 \mu - z^2$$

Which can be solved for $\sinh^2 \mu$ using the quadratic formula:

$$\sinh^2 \mu = \frac{-r_p^2 + r_c^2 + z^2 + \sqrt{(r_p^2 - r_c^2 - z^2)^2 + 4 r_p^2 z^2}}{2 r_p^2} \tag{S13}$$

An expression for $\mu_{lim}(z)$, the value of $\mu$ delimiting the integration domains, is found:

$$\mu_{lim}(z) = \sinh^{-1}\left(\frac{z}{r_p}\sqrt{\frac{1}{2}\left(1 - \frac{r_p^2 - r_c^2}{z^2} + \sqrt{\left(\frac{r_p^2 - r_c^2}{z^2} - 1\right)^2 + 4\frac{r_p^2}{z^2}}\right)}\right) = \sinh^{-1}\left(\frac{z}{r_p}\xi(z)\right) \tag{S14}$$

First consider the $R'^{bot}_{cyl}$ contribution, corresponding to the region where oblate spheroids at $\mu$ intersect solely with the bottom extremity of the cylindrical obstruction. From the coordinate transforms (Eq. S1), the fixed-$z$ horizontal bottom extremity is parametrized as:

$$\sin \nu = \frac{z_o - \ell_c}{r_p \sinh \mu} \tag{S15}$$

The resistance for the domain accounting for the flat bottom of the cylindrical obstruction is thus:

$$R'^{bot}_{cyl} = \frac{1}{2\sigma\pi r_p} \int_{\sinh^{-1}\left(\frac{z_o-\ell_c}{r_p}\right)}^{\sinh^{-1}\left(\frac{z_o-\ell_c}{r_p}\xi(z_o-\ell_c)\right)} \frac{\operatorname{sech}\mu \, d\mu}{\sin \nu}$$

$$= \frac{1}{2\sigma\pi(z_o - \ell_c)} \int_{\sinh^{-1}\left(\frac{z_o-\ell_c}{r_p}\right)}^{\sinh^{-1}\left(\frac{z_o-\ell_c}{r_p}\xi(z_o-\ell_c)\right)} \frac{\sinh\mu \, d\mu}{\cosh\mu}$$

$$= \frac{1}{2\sigma\pi(z_o - \ell_c)} [\ln(\cosh\mu)]_{\sinh^{-1}\left(\frac{z_o-\ell_c}{r_p}\right)}^{\sinh^{-1}\left(\frac{z_o-\ell_c}{r_p}\xi(z_o-\ell_c)\right)}$$

$$R'^{bot}_{cyl} = \frac{1}{4\sigma\pi(z_o - \ell_c)} \ln\left(\frac{1 + \left(\frac{z_o - \ell_c}{r_p}\xi(z_o - \ell_c)\right)^2}{1 + \left(\frac{z_o - \ell_c}{r_p}\right)^2}\right) \tag{S16}$$

Now consider the $R'^{side}_{cyl}$ contribution, corresponding to the region where constant-$\mu$ surfaces intersect only with the side of the cylinder, which is trivially parametrized by $\rho_o(\mu) = r_c$, as per the main text, and delimited by $\mu = \sinh^{-1}\left(\frac{z_o - \ell_c}{r_p}\xi(z_o - \ell_c)\right)$ and $\mu = \sinh^{-1}\left(\frac{z_o + \ell_c}{r_p}\right)$:

$$R'^{side}_{cyl} = \frac{1}{2\pi\sigma r_p} \int_{\sinh^{-1}\left(\frac{z_o-\ell_c}{r_p}\xi(z_o-\ell_c)\right)}^{\sinh^{-1}\left(\frac{z_o+\ell_c}{r_p}\right)} \frac{\text{sech}\,\mu\, d\mu}{\sqrt{1 - \frac{r_c^2}{r_p^2}\text{sech}^2\mu}}$$

$$= \frac{1}{2\pi\sigma r_p} \int_{\frac{\pi}{2}-\tan^{-1}\left(\frac{z_o+\ell_c}{r_p}\right)}^{\frac{\pi}{2}-\tan^{-1}\left(\frac{z_o-\ell_c}{r_p}\xi(z_o-\ell_c)\right)} \frac{d\theta}{\sqrt{1 - \frac{r_c^2}{r_p^2}\sin^2\theta}}$$

$$= \frac{1}{2\sigma r_p}\left[\frac{1}{\pi}F\left(\frac{\pi}{2} - \tan^{-1}\left(\frac{z_o - \ell_c}{r_p}\xi(z_o - \ell_c)\right), \frac{r_c}{r_p}\right) - \frac{1}{\pi}F\left(\frac{\pi}{2} - \tan^{-1}\left(\frac{z_o + \ell_c}{r_p}\right), \frac{r_c}{r_p}\right)\right] \quad (S17)$$

$R'^{side-top}_{cyl}$, the last obstructed contribution, accounts for the region where oblate spheroidal slices intersect both with the side of the cylinder and its top (Fig. S2). Note, as a result, that the $\nu$ integral in the $dR$ calculation thus must be split into two integration domains:

$$R'^{side-top}_{cyl} = \frac{1}{2\pi\sigma r_p} \int_{\sinh^{-1}\left(\frac{z_o+\ell_c}{r_p}\right)}^{\sinh^{-1}\left(\frac{z_o+\ell_c}{r_p}\xi(z_o+\ell_c)\right)} \frac{\text{sech}\,\mu\, d\mu}{\int_0^{\nu_{side}}\cos\nu\, d\nu + \int_{\nu_{top}}^{\frac{\pi}{2}}\cos\nu\, d\nu}$$

$$= \frac{1}{2\pi\sigma r_p} \int_{\sinh^{-1}\left(\frac{z_o+\ell_c}{r_p}\right)}^{\sinh^{-1}\left(\frac{z_o+\ell_c}{r_p}\xi(z_o+\ell_c)\right)} \frac{\text{sech}\,\mu\, d\mu}{1 - \frac{z_o + \ell_c}{r_p \sinh\mu} + \sqrt{1 - \frac{r_c^2}{r_p^2}\text{sech}^2\mu}} \quad (S18)$$

Lastly, the resistance of the unobstructed regions is calculated as:

$$R_{free} = R^{bot}_{free} + R^{top}_{free}$$

$$= \frac{1}{2\pi\sigma r_p} \int_{-\infty}^{\sinh^{-1}\left(\frac{z_o-\ell_c}{r_p}\right)} \text{sech}\,\mu\, d\mu + \frac{1}{2\pi\sigma r_p} \int_{\sinh^{-1}\left(\frac{z_o+\ell_c}{r_p}\xi(z_o+\ell_c)\right)}^{\infty} \text{sech}\,\mu\, d\mu$$

$$= \frac{1}{2\pi\sigma r_p}\left(\tan^{-1}\left(\frac{z_o - \ell_c}{r_p}\right) + \frac{\pi}{2}\right) + \frac{1}{2\pi\sigma r_p}\left(\frac{\pi}{2} - \tan^{-1}\left(\frac{z_o + \ell_c}{r_p}\xi(z_o + \ell_c)\right)\right)$$

$$= \frac{1}{2\sigma r_p}\left[1 + \frac{1}{\pi}\tan^{-1}\left(\frac{z_o - \ell_c}{r_p}\right) - \frac{1}{\pi}\tan^{-1}\left(\frac{z_o + \ell_c}{r_p}\xi(z_o + \ell_c)\right)\right] \quad (S19)$$

<u>Inside the pore ($|z_o| \leq \ell_c$)</u>

Again, as per Figure S1 and similarly to Equation S11, the resistances of five separate domains need to be considered when calculating the conductance of flat-ended obstructions bridging across the membrane:

$$G_{cyl}^b(r_c, \ell_c, z_o) = \left[R_{free}^{bot} + R_{cyl}^{'bot-side} + R_{cyl}^{'side} + R_{cyl}^{'side-top} + R_{free}^{top}\right]^{-1} \quad (S20)$$

Similarly to the $|z_o| > \ell_c$ case, the resistance of the unobstructed regions is found to be:

$$R_{free}^{bot} + R_{free}^{top} = \frac{1}{2\pi\sigma r_p} \int_{-\infty}^{\sinh^{-1}\left(\frac{z_o-\ell_c}{r_p}\xi(z_o-\ell_c)\right)} \operatorname{sech}\mu \, d\mu + \frac{1}{2\pi\sigma r_p} \int_{\sinh^{-1}\left(\frac{z_o+\ell_c}{r_p}\xi(z_o+\ell_c)\right)}^{\infty} \operatorname{sech}\mu \, d\mu$$

$$= \frac{1}{2\sigma r_p}\left[1 + \frac{1}{\pi}\tan^{-1}\left(\frac{z_o-\ell_c}{r_p}\xi(z_o-\ell_c)\right) - \frac{1}{\pi}\tan^{-1}\left(\frac{z_o+\ell_c}{r_p}\xi(z_o+\ell_c)\right)\right] \quad (S21)$$

The $R_{cyl}^{'bot-side}$ terms accounts for the segment where oblate spheroidal slices intersect both with the side of the cylinder and its bottom (Fig. S2). As a result, the $v$ integral in the $dR$ calculation must again be split into two integration domains:

$$R_{cyl}^{'bot-side} = \frac{1}{2\pi\sigma r_p} \int_{\sinh^{-1}\left(\frac{z_o-\ell_c}{r_p}\xi(z_o-\ell_c)\right)}^{\sinh^{-1}\left(\frac{z_o-\ell_c}{r_p}\right)} \frac{\operatorname{sech}\mu \, d\mu}{[\sin v]_0^{v_{side}} + [\sin v]_{v_{bot}}^{\frac{\pi}{2}}}$$

$$= \frac{1}{2\pi\sigma r_p} \int_{\sinh^{-1}\left(\frac{z_o-\ell_c}{r_p}\xi(z_o-\ell_c)\right)}^{\sinh^{-1}\left(\frac{z_o-\ell_c}{r_p}\right)} \frac{\operatorname{sech}\mu \, d\mu}{1 - \frac{z_o-\ell_c}{r_p \sinh\mu} + \sqrt{1 - \frac{r_c^2}{r_p^2}\operatorname{sech}^2\mu}} \quad (S22)$$

The $R_{cyl}^{'side}$ terms accounts for the segment where oblate spheroidal slices intersect only with the side of the cylinder (Fig. S2), and is delimited by $\mu_{\pm} = \sinh^{-1}\left(\frac{z_o \pm \ell_c}{r_p}\right)$:

$$R_{cyl}^{'side} = \frac{1}{2\pi\sigma r_p} \int_{\sinh^{-1}\left(\frac{z_o-\ell_c}{r_p}\right)}^{\sinh^{-1}\left(\frac{z_o+\ell_c}{r_p}\right)} \frac{\operatorname{sech}\mu \, d\mu}{\sqrt{1 - \frac{r_c^2}{r_p^2}\operatorname{sech}^2\mu}}$$

$$= \frac{1}{2\sigma r_p}\left[\frac{1}{\pi}F\left(\frac{\pi}{2} - \tan^{-1}\left(\frac{z_o-\ell_c}{r_p}\right)\right) - \frac{1}{\pi}F\left(\frac{\pi}{2} - \tan^{-1}\left(\frac{z_o+\ell_c}{r_p}\right)\right)\right] \quad (S23)$$

The $R_{cyl}^{'side-top}$ terms accounts for the segment where oblate spheroidal slices intersect both with the side of the cylinder and its top (Fig. S2):

$$R_{cyl}^{'side-top} = \frac{1}{2\pi\sigma r_p} \int_{\sinh^{-1}\left(\frac{z_o+\ell_c}{r_p}\right)}^{\sinh^{-1}\left(\frac{z_o+\ell_c}{r_p}\xi(z_o+\ell_c)\right)} \frac{\operatorname{sech}\mu \, d\mu}{[\sin v]_0^{v_{side}} + [\sin v]_{v_{top}}^{\frac{\pi}{2}}}$$

$$= \frac{1}{2\pi\sigma r_p} \int_{\sinh^{-1}\left(\frac{z_o+\ell_c}{r_p}\right)}^{\sinh^{-1}\left(\frac{z_o+\ell_c}{r_p}\xi(z_o+\ell_c)\right)} \frac{\operatorname{sech}\mu \, d\mu}{1 - \frac{z_o+\ell_c}{r_p \sinh\mu} + \sqrt{1 - \frac{r_c^2}{r_p^2}\operatorname{sech}^2\mu}} \quad (S24)$$

We note that the above equations do not apply for very thin and wide cylinders whose treatment requires the consideration of an oblate spheroidal segment in which constant-$\mu$ surfaces intersect both the bottom and top surfaces of the cylinders.

Figure S3 plots Equation S11 for cylinders of length $L_c = 10 r_p$ with radii smaller and larger than the pore radius. For $r_c \geq r_p$, the conductance is zero when $z_o \leq L_c/2$, as expected.

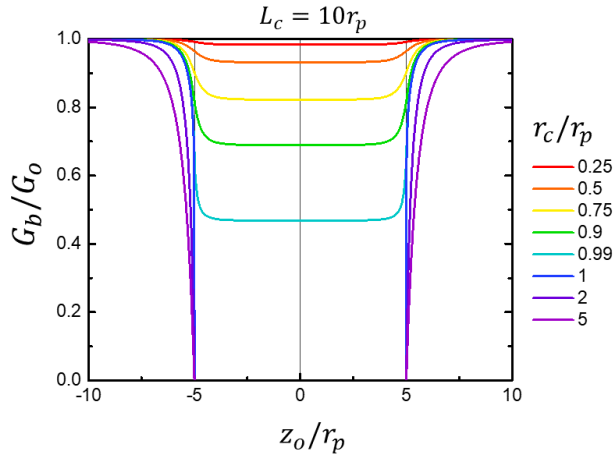

**Figure S3.** $z$-dependence of 2D pore conductance in the presence of cylinders of different radii and fixed length $L_c = 10 r_p$.

### S4. Spherical Obstruction

Here the full derivation of a spherical obstruction in a 2D pore is presented.

<u>Centered case ($z_o = 0$)</u>

The equation describing the surface of the spherical obstruction is $r_s^2 = \rho^2 + z^2$, from which an expression for $z^2 = r_s^2 - \rho^2$ is obtained and inserted into the equation of a constant-$\mu$ oblate spheroid:

$$1 = \frac{\rho^2}{r_p^2 \cosh^2 \mu} + \frac{r_s^2 - \rho^2}{r_p^2 \sinh^2 \mu} \quad (S25)$$

Through simple manipulations, an expression for the $\rho$-parametrization can be obtained:

$$\rho_o^2 = \left(r_s^2 - r_p^2 \sinh^2 \mu\right) \cosh^2 \mu \quad (S26)$$

The resistance of the oblate spheroidal segment containing the spherical obstruction is thus:

$$R'_{sph}(r_s) = \frac{1}{2\pi\sigma r_p} \int_{\sinh^{-1}\left(\frac{-r_s}{r_p}\right)}^{\sinh^{-1}\left(\frac{r_s}{r_p}\right)} \frac{\text{sech}\,\mu\,d\mu}{\sqrt{1 - \frac{r_s^2 - r_p^2 \sinh^2 \mu}{r_p^2}}}$$

$$= \frac{1}{\pi\sigma r_p} \int_{0}^{\sinh^{-1}\left(\frac{r_s}{r_p}\right)} \frac{\text{sech}^2 \mu\,d\mu}{\sqrt{1 - \frac{r_s^2}{r_p^2}}\sqrt{1 + \frac{r_s^2/r_p^2}{1 - \frac{r_s^2}{r_p^2}} \tanh^2 \mu}}$$

$$= \frac{1}{\pi\sigma r_s} \left[ \sinh^{-1}\left( \frac{r_s/r_p}{\sqrt{1 - r_s^2/r_p^2}} \tanh(\mu) \right) \right]_{0}^{\sinh^{-1}\left(\frac{r_s}{r_p}\right)}$$

$$= \frac{1}{\pi\sigma r_s} \left( \sinh^{-1}\left( \frac{r_s^2/r_p^2}{\sqrt{1 - (r_s^2/r_p^2)^2}} \right) \right)$$

$$\Rightarrow R'_{sph}(r_s) = \frac{1}{\pi\sigma r_s} \tanh^{-1}\left( \frac{r_s^2}{r_p^2} \right) \tag{S27}$$

By including the contribution from the unobstructed segments (Equation 6, main text), the conductance of a 2D nanopore in the presence of a centered spherical obstruction is:

$$\frac{G_b^{sph}}{G_o}(r_s) = \left[ 1 - \frac{2}{\pi} \tan^{-1}\left(\frac{r_s}{r_p}\right) + \frac{2\,r_p}{\pi\,r_s} \tanh^{-1}\left(\frac{r_s^2}{r_p^2}\right) \right]^{-1} \tag{S28}$$

Non-Centered case ($z_o \neq 0$)

The surface of the sphere translated along the z axis by $z_o$ is described by the equation:

$$\rho^2 + (z - z_o)^2 = r_s^2 \tag{S29}$$

From Equation S29, an expression for $z^2$ can be obtained, and inserted into the constant-$\mu$ oblate spheroidal equation:

$$\frac{\rho^2}{r_p^2 \cosh^2 \mu} + \frac{r_s^2 - \rho^2 + 2z_o\sqrt{r_s^2 - \rho^2} + z_o^2}{r_p^2 \sinh^2 \mu} = 1 \tag{S30}$$

Equation S30 can be rewritten as a quadratic equation with respect to $\rho^2 \text{sech}^2 \mu$:

$$0 = \rho^4 \text{sech}^4 \mu - \left(2(r_s^2 + z_0^2 - r_p^2 \sinh^2 \mu) - 4z_0^2 \cosh^2 \mu\right)\rho^2 \text{sech}^2 \mu$$
$$+ \left(\left(r_s^2 + z_0^2 - r_p^2 \sinh^2 \mu\right)^2 - 4z_0^2 r_s^2\right) \tag{S31}$$

Inserting the positive quadratic formula solution for $\rho^2 \operatorname{sech}^2 \mu$ into Equation 5 of the main text, i.e. $dR = (2\pi\sigma r_p)^{-1}/\sqrt{1 - \rho^2 \operatorname{sech}^2 \mu / r_p^2}$, results in an expression for the resistance of the obstructed oblate spheroidal segment:

$$\frac{R'_{sph}}{R_o}(r_s, z_o) = \frac{1}{\pi} \int_{\sinh^{-1}\left(\frac{z_o - r_s}{r_p}\right)}^{\sinh^{-1}\left(\frac{z_o + r_s}{r_p}\right)} \frac{\operatorname{sech}^2 \mu \, d\mu}{\sqrt{1 - \frac{r_s^2 + z_o^2}{r_p^2}\operatorname{sech}^2 \mu + 2\frac{z_o^2}{r_p^2} - 2\frac{z_o}{r_p}\tanh\mu \sqrt{1 + \frac{z_o^2}{r_p^2} - \frac{r_s^2}{r_p^2}\operatorname{sech}^2 \mu}}} \quad (S32)$$

Here, $R_o = 1/2\sigma r_p$ is the resistance of an open 2D pore. The contributions from the two unobstructed segments can above and below the sphere can also be considered (as before, e.g. see Eq. S7):

$$R_{free}(r_s, z_o) = R_{free}^{bot} + R_{free}^{top}$$

$$= \frac{1}{2\pi\sigma r_p} \int_{-\infty}^{\sinh^{-1}\left(\frac{z_o - r_s}{r_p}\right)} \operatorname{sech} \mu \, d\mu + \frac{1}{2\pi\sigma r_p} \int_{\sinh^{-1}\left(\frac{z_o + r_s}{r_p}\right)}^{\infty} \operatorname{sech} \mu \, d\mu$$

$$= \frac{1}{2\sigma r_p}\left(1 + \frac{1}{\pi}\tan^{-1}\left(\frac{z_o - r_s}{r_p}\right) - \frac{1}{\pi}\tan^{-1}\left(\frac{z_o + r_s}{r_p}\right)\right) \quad (S33)$$

The general expression for the conductance of a 2D pore in the presence of a spherical obstruction of radius $r_s$ at a distance $z_o$ from the pore center is thus:

$$\left(\frac{G_b^{sph}}{G_o}\right)^{-1} = 1 + \frac{1}{\pi}\tan^{-1}\left(\frac{z_o - r_s}{r_p}\right) - \frac{1}{\pi}\tan^{-1}\left(\frac{z_o + r_s}{r_p}\right)$$

$$+ \frac{1}{\pi}\int_{\sinh^{-1}\left(\frac{z_o - r_s}{r_p}\right)}^{\sinh^{-1}\left(\frac{z_o + r_s}{r_p}\right)} \frac{\operatorname{sech}^2 \mu \, d\mu}{\sqrt{1 - \frac{r_s^2 + z_o^2}{r_p^2}\operatorname{sech}^2 \mu + 2\frac{z_o^2}{r_p^2} - 2\frac{z_o}{r_p}\tanh\mu \sqrt{1 + \frac{z_o^2}{r_p^2} - \frac{r_s^2}{r_p^2}\operatorname{sech}^2 \mu}}} \quad (S34)$$

Equation S34 is plotted in Figure 2c of the main text alongside conductance values from corresponding finite element simulations. Interestingly, we note that unlike with cylinders, no special treatment is required for spheres bigger than the pores, as demonstrated in Figure S4 which plots examples of Equation S34 for both $r_s < r_p$ and $r_s \geq r_p$.

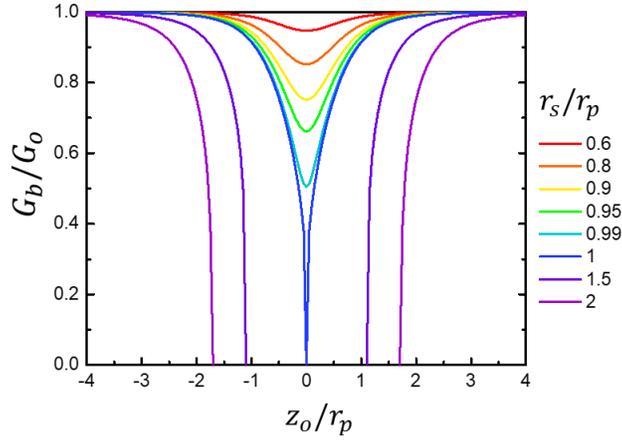

**Figure S4.** $z$-dependence of the conductance of 2D pores in the presence of spherical obstructions.

## S5. Fractional Blockage Error Plot

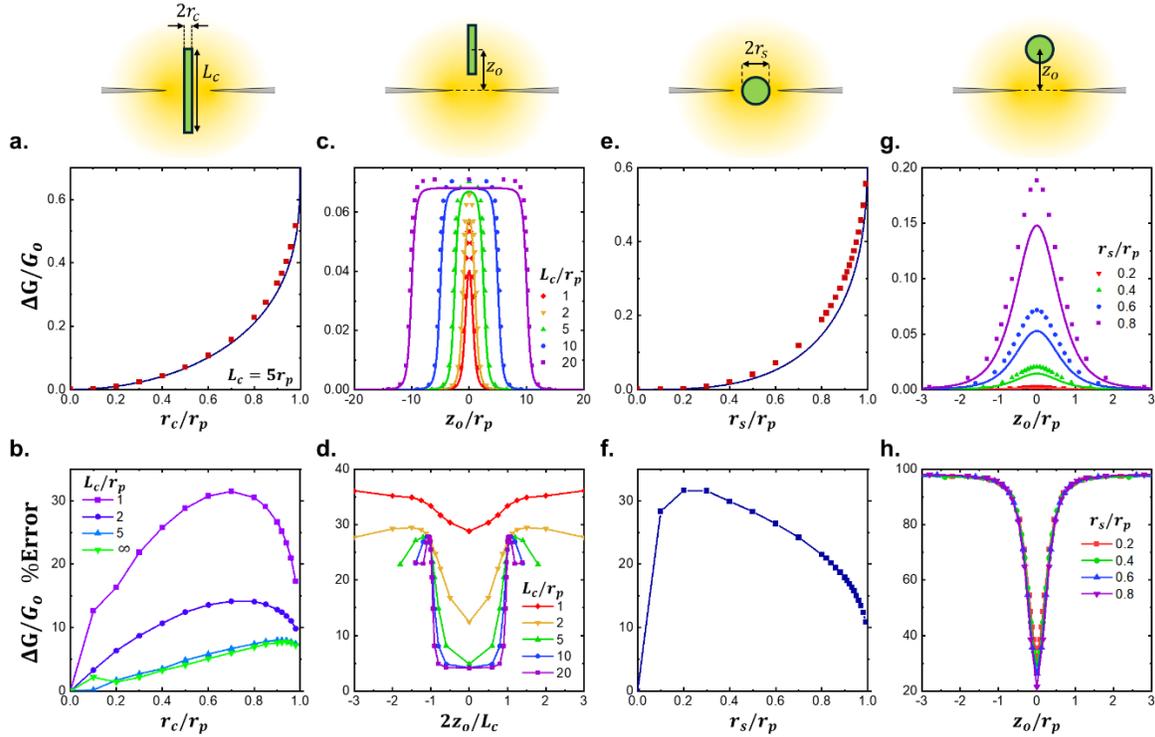

**Figure S5.** Dependence of normalized conductance blockage $\Delta G/G_o$ on obstruction dimensions and distance from a 2D pore. Lines are from model predictions, whereas points are from finite element simulations. **a)** $\Delta G/G_o$ vs cylindrical obstruction radius $r_c$. **b)** Absolute error percentage between modelled and simulated $\Delta G/G_o$ values from (a). **c)** $\Delta G/G_o$ vs $z_o$ for cylindrical obstruction of radius $r_c = 0.5 r_p$ and different lengths $L_p$. **d)** Absolute error percentage between modelled and simulated $\Delta G/G_o$ values from (c). **e)** $\Delta G/G_o$ vs centered ($z_o = 0$) spherical obstruction radius $r_s$. **f)** Absolute error percentage between modelled and simulated $\Delta G/G_o$ values from (e). **g)** $\Delta G/G_o$ vs $z_o$ for spherical obstructions of different radii. **h)** Absolute error percentage between modelled and simulated $\Delta G/G_o$ values from (g).

## S6. Rotationally Symmetric Obstructions – More Examples

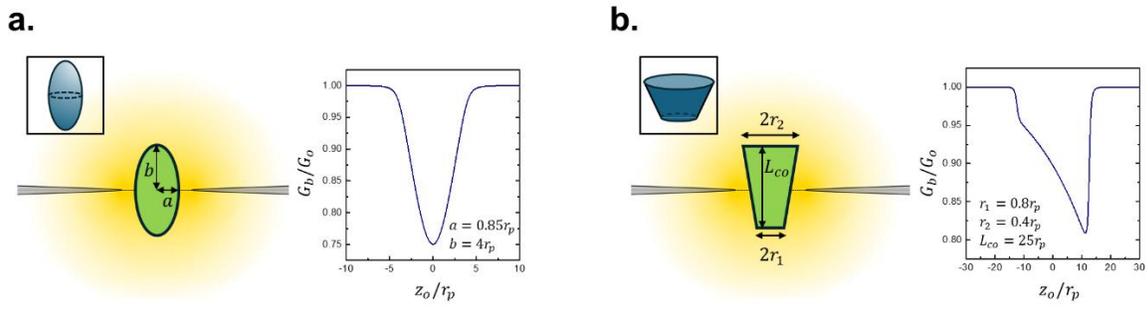

**Figure S6.** Conductance of a 2D pore in the presence of an obstruction with **(a)** a spheroidal geometry and **(b)** with a truncated conical geometry.

### Spheroid Centered Inside a 2D Pore

The surface of a spheroid at a distance $z_o$ from a 2D pore with dimensions $a$ along the $\rho$-axis and $b$ along the $z$-axis is described by the equation:

$$\frac{\rho^2}{a^2} + \frac{(z-z_o)^2}{b^2} = 1 \tag{S35}$$

The $\rho_o(\mu)$ parameterization of the spheroid obstruction is found by isolating for $z^2$ in Equation S35 and inserting the expression into the equation of a constant-$\mu$ surface, resulting in (for $z_o$ = 0):

$$\rho_o^2(\mu) = \frac{\sinh^2\mu - \dfrac{b^2}{r_p^2}}{\sinh^2\mu - \dfrac{b^2}{a^2}\cosh^2\mu} r_p^2 \cosh^2\mu \tag{S36}$$

The corresponding expression for the resistance of the oblate spheroidal segment containing the spheroidal obstruction is thus:

$$R'_{ellipse}(a,b) = \frac{1}{2\pi\sigma r_p} \int_{\sinh^{-1}\left(-\frac{b}{r_p}\right)}^{\sinh^{-1}\left(\frac{b}{r_p}\right)} \frac{\operatorname{sech}\mu\, d\mu}{\sqrt{1 - \dfrac{\sinh^2\mu - \dfrac{b^2}{r_p^2}}{\sinh^2\mu - \dfrac{b^2}{a^2}\cosh^2\mu}}}$$

$$= \frac{1}{2\pi\sigma r_p} \int_{\sinh^{-1}\left(-\frac{b}{r_p}\right)}^{\sinh^{-1}\left(\frac{b}{r_p}\right)} \frac{\sqrt{1 - \dfrac{a^2}{b^2}\tanh^2\mu}}{\sqrt{1 - \dfrac{a^2}{r_p^2}\operatorname{sech}^2\mu}} \operatorname{sech}\mu\, d\mu \tag{S37}$$

Equation S37 presents no closed form solution, however, note that in the two limits of $a = b$ and $b \to \infty$, $R'_{ellipse}(\mu)$ corresponds exactly to the expressions for the spherical obstruction (Equation S27), and the infinitely long cylindrical obstruction (Equation S8), respectively. The resistance of a nanopore obstructed by a centered spheroid is therefore:

$$\frac{G_b^{ellipse}}{G_o}(a,b) = \left[1 - \frac{2}{\pi}\tan^{-1}\left(\frac{b}{r_p}\right) + \frac{1}{\pi}\int_{\sinh^{-1}\left(-\frac{b}{r_p}\right)}^{\sinh^{-1}\left(\frac{b}{r_p}\right)} \frac{\sqrt{1 - \frac{a^2}{b^2}\tanh^2\mu}}{\sqrt{1 - \frac{a^2}{r_p^2}\text{sech}^2\mu}}\,\text{sech}\,\mu\,d\mu\right]^{-1} \quad (S38)$$

### Spheroid Centered at distance $z_o$ from a 2D Pore

Let an spheroid of semi-axes $a$ and $b$ along the $\rho$ and $z$ dimensions, respectively, have its center be at a distance $z_o$ from a 2D nanopore. Its surface can be parametrized by finding its intersection points with constant-$\mu$ ellipses, as was done above. Doing so results in the following expression for the resistance of a punctured ellipsoidal slice:

$$\frac{R'_{ellipse}}{R_o} = 2\sigma r_p \int dR(\mu)$$
$$= \frac{1}{\pi}\int_{\sinh^{-1}\left(\frac{z_o-b}{r_p}\right)}^{\sinh^{-1}\left(\frac{z_o+b}{r_p}\right)} \frac{\text{sech}\,\mu\,d\mu}{\sqrt{1 + \frac{\left(\tanh^2\mu - \frac{b^2}{a^2}\right)\left(\frac{b^2+z_o^2}{r_p^2} - \sinh^2\mu\right) + 2\frac{z_o^2 b^2}{r_p^2 a^2} - \text{sign}(\mu)2\frac{z_o b}{r_p a}\sqrt{\left(\tanh^2\mu - \frac{b^2}{a^2}\right)\left(\frac{b^2+z_o^2}{r_p^2} - \sinh^2\mu\right) + \frac{z_o^2 b^2}{r_p^2 a^2} + \frac{a^2}{r_p^2}\left(\tanh^2\mu - \frac{b^2}{a^2}\right)^2}}{\left(\tanh^2\mu - \frac{b^2}{a^2}\right)^2\cosh^2\mu}}} \quad (S39)$$

The total resistance in the system can be expressed as:

$$R_b^{ellipse}(a,b,z_o) = R_{free}(-\infty, z_o - b) + R'_{ellipse} + R_{free}(z_o + b, \infty) \quad (S40)$$

and the normalized blocked-state conductance can be expressed as:

$$\frac{G_b^{ellipse}}{G_o}(a,b,z_o) = \left[1 + \frac{1}{\pi}\tan^{-1}\left(\frac{z_o - b}{r_p}\right) - \frac{1}{\pi}\tan^{-1}\left(\frac{z_o + b}{r_p}\right) + \frac{R'_{ellipse}}{R_o}\right]^{-1} \quad (S41)$$

The z dependence of Equation S41 is plotted in Figure S5a for an spheroid of dimension $a = 0.85 r_p$ and $b = 4 r_p$.

### Truncated Cone

Here we consider a truncated cone of length $L_{co} = 2\ell_{co}$, centered at $z_o$ with one of its faces located at $z = z_o - \ell_c$ having radius $r_1$ and the other located at $z = z_o + \ell_{co}$ with radius $r_2$. The surface of the conical section is parametrized by the following equation:

$$\rho_o(z) = \frac{r_2 - r_1}{2\ell_{co}}(z - z_o) + \frac{r_1 + r_2}{2} \quad (S42)$$

The parametrization of $\rho_o$ with respect to $\mu$ can be found by isolating $z$ from the above and inserting this result into the expression of a constant-$\mu$ surface, as before. The infinitesimal resistance of an oblate spheroidal slice containing the obstruction can then be found using Equation 5 in the main text:

$$dR'_{co}(\mu) = \frac{1}{2\pi\sigma r_p}\frac{\text{sech}\,\mu\,d\mu}{\sqrt{1 - \frac{\left(\frac{2\ell_{co}}{r_2-r_1}\left(\ell_{co}\frac{r_2+r_1}{r_2-r_1} - z_o\right) + \text{sign}(\mu(r_2-r_1))\sqrt{\left(\tanh^2\mu + \frac{4\ell_{co}^2}{(r_2-r_1)^2}\right)r_p^2\sinh^2\mu - \tanh^2\mu\left(z_o - \ell_{co}\frac{r_2+r_1}{r_2-r_1}\right)^2}\right)}{r_p^2\cosh^2\mu\left(\tanh^2\mu + \frac{4\ell_{co}^2}{(r_2-r_1)^2}\right)^2}}} \quad (S43)$$

The total resistance in the system can be expressed as:

$$R_b^{co}(r_1, r_2, \ell_{co}, z_o) = R_{free}(-\infty, z_o - \ell_{co}) + \int_{\sinh^{-1}\frac{z_o-\ell_{co}}{r_p}}^{\sinh^{-1}\frac{z_o+\ell_{co}}{r_p}} dR'_{co}(\mu) + R_{free}(z_o + \ell_{co}, \infty) \quad \text{(S44)}$$

and the normalized blocked-state conductance as:

$$\frac{G_b^{co}}{G_o} = \left[1 + \frac{1}{\pi}\tan^{-1}\left(\frac{z_o - \ell_{co}}{r_p}\right) - \frac{1}{\pi}\tan^{-1}\left(\frac{z_o + \ell_{co}}{r_p}\right) + 2\sigma r_p \int_{\sinh^{-1}\frac{z_o-\ell_{co}}{r_p}}^{\sinh^{-1}\frac{z_o+\ell_{co}}{r_p}} dR'_{co}(\mu)\right]^{-1} \quad \text{(S45)}$$

The z dependence of Equation S45 is plotted in Figure S5b for a truncated cone of dimension $r_1 = 0.8r_p$, $r_2 = 0.4r_p$ and $L_{co} = 2\ell_{co} = 25r_p$.

### S7. Rotationally Asymmetric Obstructions – Wedged Cylinder

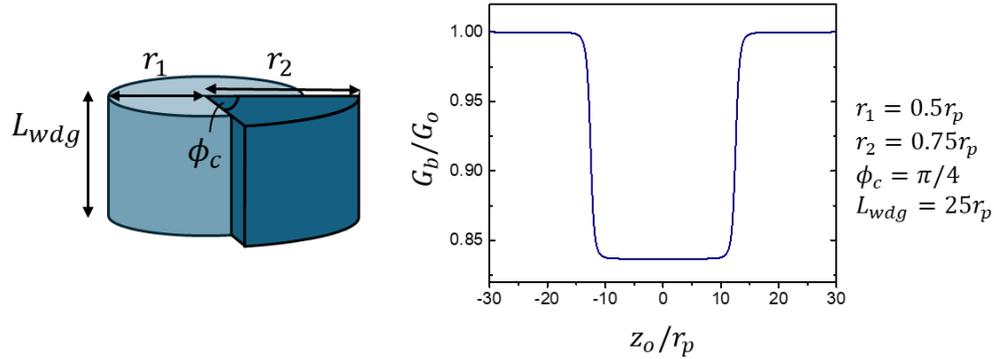

**Figure S7.** Conductance of a 2D pore in the presence of a wedged cylindrical obstruction.

Consider the simple rotationally asymmetric case of a wedged cylinder of length $L_{wdg} = 2\ell_{wdg}$, whose surface is defined as:

$$\rho_o(\phi) = \begin{cases} r_1, & \text{for } \phi_c/2 < \phi < 2\pi - \phi_c/2 \\ r_2, & \text{for } -\phi_c/2 < \phi < \phi_c/2 \end{cases} \quad \text{(S46)}$$

The resistance of the punctured oblate spheroidal slice at $\mu$ can be calculated as:

$$dR'_{wedge}(\mu) = \frac{1}{\sigma r_p} \frac{\text{sech}\,\mu}{\int_{\phi_c/2}^{2\pi-\phi_c/2}\sqrt{1 - \frac{r_1^2}{r_p^2}\text{sech}^2\mu}\,d\phi + \int_{-\phi_c/2}^{\phi_c/2}\sqrt{1 - \frac{r_2^2}{r_p^2}\text{sech}^2\mu}\,d\phi}$$

$$= \frac{1}{2\sigma\pi r_p} \frac{\text{sech}\,\mu}{\left(1 - \frac{\phi_c}{2\pi}\right)\sqrt{1 - \frac{r_1^2}{r_p^2}\text{sech}^2\mu} + \frac{\phi_c}{2\pi}\sqrt{1 - \frac{r_2^2}{r_p^2}\text{sech}^2\mu}} \quad \text{(S47)}$$

The total resistance in the system can be expressed as:

$$R_b^{wedge}(r_1, r_2, \phi_c, \ell_{wdg}, z_o) = R_{free}(-\infty, z_o - \ell_{wdg}) + R'_{wedge} + R_{free}(z_o + \ell_{wdg}, \infty) \quad \text{(S48)}$$

and the normalized blocked-state conductance can be expressed as:

$$\frac{G_b^{wedge}}{G_o} = \left[1 + \frac{1}{\pi}\tan^{-1}\left(\frac{z_o - \ell_w}{r_p}\right) - \frac{1}{\pi}\tan^{-1}\left(\frac{z_o + \ell_w}{r_p}\right) + 2\sigma r_p \int_{\sinh^{-1}\frac{z_o-\ell_w}{r_p}}^{\sinh^{-1}\frac{z_o+\ell_w}{r_p}} dR'_{wedge}(\mu)\right]^{-1} \quad (S49)$$

## S8. Rotationally Asymmetric Obstructions – Off-Axis Effects

### Off-Centered Cylinder

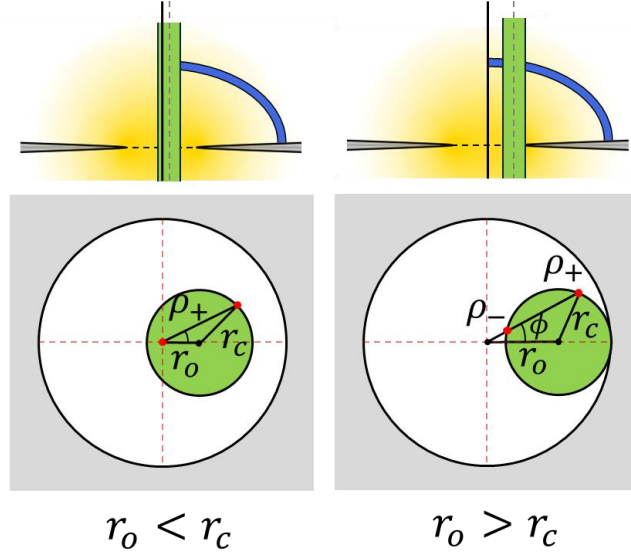

$r_o < r_c \qquad r_o > r_c$

**Figure S8**. Cross-sectional view of the x-y plane showing the geometrical parameters of a 2D pore in the presence of an infinitely long cylindrical obstruction at a distance $r_o$ from the pore center.

Consider an infinitely long cylindrical obstruction of radius $r_c$, lying parallel to the z axis, but centred at a distance $r_o$ away from the central pore axis, as per Figure S7. The parametrization of the obstruction's surface is found by calculating the distance $\rho_o$ at which the constant-$\phi$ planes and the cylindrical obstruction intersect. As per Figure S7, this can be achieved using the trigonometry of the system cross-section in the $z = 0$ plane:

$$\rho_\pm(\phi) = r_o \cos\phi \pm \sqrt{r_c^2 - r_o^2 \sin^2\phi} \quad (S50)$$

Note that the scenarios for which $r_o < r_c$ and $r_o > r_c$ need to be treated separately. For $r_o < r_c$, the resistance of the obstructed system is found to be:

$$R'_{cyl}(r_o < r_c) = \frac{1}{\sigma r_p} \int_{-\infty}^{\infty} \frac{\text{sech}\,\mu\, d\mu}{\int_0^{2\pi}[\sin v]_0^{v_o(\phi)} d\phi}$$

$$= \frac{1}{\sigma r_p} \int_{-\infty}^{\infty} \frac{\text{sech}\,\mu\, d\mu}{\int_0^{2\pi}\sqrt{1 - \frac{\rho_+^2(\phi)}{r_p^2}\text{sech}^2\,\mu}\, d\phi}$$

$$= \frac{1}{\sigma r_p} \int_{-\infty}^{\infty} \frac{\operatorname{sech} \mu \, d\mu}{\int_0^{2\pi} \sqrt{1 - \left(\frac{r_o}{r_p} \cos \phi + \sqrt{\frac{r_c^2}{r_p^2} - \frac{r_o^2}{r_p^2} \sin^2 \phi}\right)^2 \operatorname{sech}^2 \mu} \, d\phi} \tag{S51}$$

For $r_o > r_c$, the resistance calculation involves integrating over two distinct $v$ domains, and thereby using both the $\rho_+$ and $\rho_-$ solutions of Equation S50. Moreover, note that not all constant-$\phi$ planes intersect the cylindrical obstruction, only those within the angles $|\phi| < \phi_c = \sin^{-1}(r_c/r_o)$:

$$R'_{cyl}(r_o > r_c) = \frac{1}{\sigma r_p} \int_{-\infty}^{\infty} \frac{\operatorname{sech} \mu \, d\mu}{\int_{\phi_c}^{2\pi - \phi_c} [\sin v]_0^{\frac{\pi}{2}} d\phi + \int_{-\phi_c}^{\phi_c} \left([\sin v]_0^{v_+(\phi)} + [\sin v]_{v_-(\phi)}^{\frac{\pi}{2}}\right) d\phi}$$

$$= \frac{1}{\sigma r_p} \int_{-\infty}^{\infty} \frac{\operatorname{sech} \mu \, d\mu}{2\pi - 2\phi_c + \int_{-\phi_c}^{\phi_c} \left(\sqrt{1 - \frac{\rho_+^2}{r_p^2} \operatorname{sech}^2 \mu} + 1 - \sqrt{1 - \frac{\rho_-^2}{r_p^2} \operatorname{sech}^2 \mu}\right) d\phi}$$

$$= \frac{1}{\sigma r_p} \int_{-\infty}^{\infty} \frac{\operatorname{sech} \mu \, d\mu}{2\pi + \int_{-\sin^{-1}\left(\frac{r_c}{r_o}\right)}^{\sin^{-1}\left(\frac{r_c}{r_o}\right)} \left(\sqrt{1 - \frac{\rho_+^2}{r_p^2} \operatorname{sech}^2 \mu} - \sqrt{1 - \frac{\rho_-^2}{r_p^2} \operatorname{sech}^2 \mu}\right) d\phi} \tag{S52}$$

Equation 17 of the main text thus denotes the blocked-state conductance, i.e. inverse values of Equations S51 and S52 normalized by the open pore conductance $2\sigma r_p$. As in Figure 3b of the main text, Figure S8a plots the dependence of the normalized conductance on $r_o$ for different values of $r_c/r_p$. Figure S8b plots the extra blockage arising from $r_o > 0$ values, calculated as the percentage of extra conductance in the $r_o = 0$ case as compared to $r_o \neq 0$.

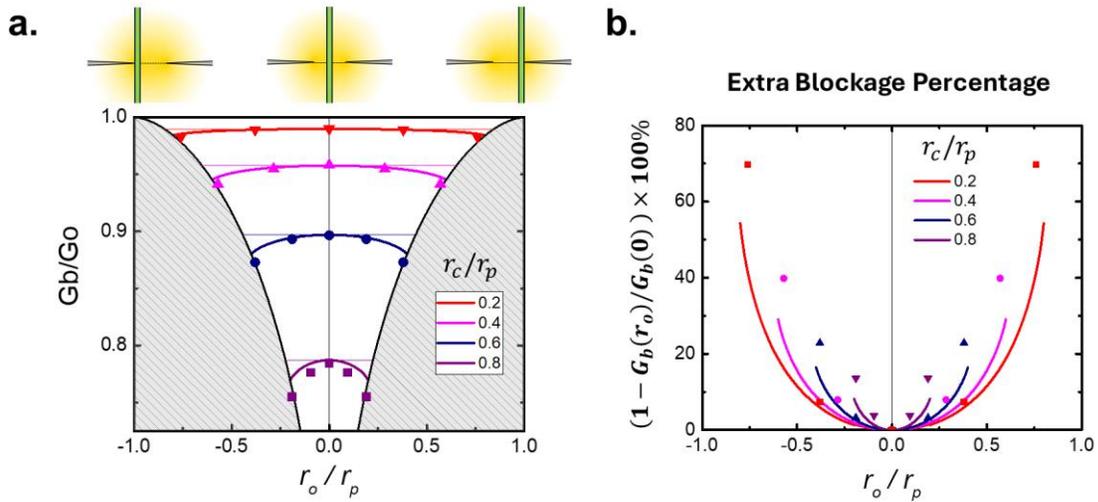

**Figure S9.** Off-axis effects for infinitely long cylindrical obstructions. **a)** Dependence of normalized blocked-state conductance $G_b^{cyl}/G_o$ on radial position of cylinder inside pore $r_o$. **b)** Extra conductance induced by centred cylinder position $r_o = 0$, compared to off-axis $r_o \neq 0$. All individual points are values calculated from finite element simulations, whereas continuous curves are from Equations S51-S52.

**Off-Centered Sphere**

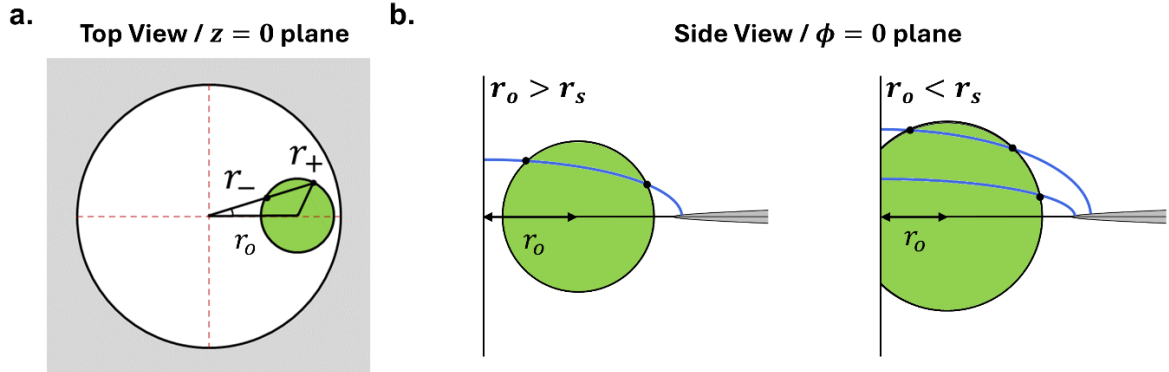

**Figure S10.** a) $z = 0$ cross-section sketch showing the simple trigonometry required to characterize the intersection circles resulting from the presence of a vertically centred spherical obstruction in a 2D pore using $r_\pm$. b) Sketch depicting the number of intersection points between the surface of a circle and an ellipse.

Unlike infinitely long cylindrical obstructions, the surface parametrization of spherical obstructions is not solely dependent on $\mu$, as is now discussed. First note that the intersection of constant-$\phi$ planes with the sphere surface is a circle. The intersection of constant-$\phi$ with constant-$\mu$ oblate spheroids is instead an ellipse. Finding the parametrization $\rho_o(\phi, \mu)$ of the spherical surface therefore relies on finding the intersection point(s) between these $\phi$-characteristic circles and ellipses.

Let us first characterize the circle resulting from the intersection of the $\phi$-plane with the vertically centred ($z_o = 0$) spherical obstruction. As shown in Figure S9a, the problem is again solved using simple trigonometry while considering the $z = 0$ plane cross-section. We can thus find $r_\pm$, defined as the intersection points of the $\phi$-plane and the sphere in the $z = 0$ plane:

$$r_\pm = r_o \cos\phi \pm \sqrt{r_s^2 - r_o^2 \sin^2\phi} \tag{S53}$$

The $\phi$-dependent radius $r_{cs}$ of the intersection circle as well as its mean radial distance from the pore center $r_{co}$ can be calculated from $r_\pm$:

$$r_{cs}(\phi) = \frac{r_+ - r_-}{2} = \sqrt{r_s^2 - r_o^2 \sin^2\phi} \tag{S54}$$

$$r_{co}(\phi) = \frac{r_- + r_+}{2} = r_o \cos\phi \tag{S55}$$

The surface of the intersection circle in the constant-$\phi$ plane is thus described by:

$$\left(\rho - r_{co}(\phi)\right)^2 + z^2 = r_{cs}^2 \tag{S56}$$

The parametrization $\rho_o(\mu, \phi)$ of the sphere surface is found by isolating $z^2$ and inserting it into the equation for the surface of a constant-$\mu$ oblate spheroid:

$$\rho_\pm(\mu, \phi) = r_{co}(\phi) \cosh^2 \mu \pm \sqrt{(r_{co}^2(\phi) - r_p^2) \cosh^2 \mu \sinh^2 \mu + r_{cs}^2 \cosh^2 \mu} \tag{S57}$$

Note that just like cylindrical obstructions, there can be one or two intersections points, as denoted by the positive and negative signs of Equation S57. Unlike for cylinders however, the number of intersection points to consider is not only dependent on $r_o$ but also on $\mu$, as shown in Figure S9b. For $r_o > r_s$ there are always two intersection points, i.e. two $\nu$ intervals over which to integrate, whereas for $r_o < r_s$, there is a value of $\mu = \mu_l$ that delimits the oblate spheroidal segments with one or two intersection points. Two intersections are considered when both $\rho_\pm$ are positive, and so $\mu_l$ is found by setting $\rho_- = 0$:

$$0 = r_o \cos \phi \cosh^2 \mu_l - \sqrt{(r_o^2 \cos^2 \phi - r_p^2) \cosh^2 \mu_l \sinh^2 \mu_l + (r_s^2 - r_o^2 \sin^2 \phi) \cosh^2 \mu_l}$$

$$0 = r_o \cos \phi \cosh \mu_l - \sqrt{(r_o^2 \cos^2 \phi - r_p^2) \sinh^2 \mu_l + (r_s^2 - r_o^2 \sin^2 \phi)}$$

$$0 = r_o^2 \cos^2 \phi \cosh^2 \mu_l - r_o^2 \cos^2 \phi \sinh^2 \mu_l + r_p^2 \sinh^2 \mu_l - r_s^2 + r_o^2 \sin^2 \phi$$

$$\sinh^2 \mu_l = \frac{r_s^2 - r_o^2}{r_p^2}$$

$$\Rightarrow \mu_l = \sinh^{-1}\left(\sqrt{\frac{r_s^2 - r_o^2}{r_p^2}}\right) \tag{S58}$$

Note that $\mu_l$ is independent of $\phi$, an important requirement for the slice method to function properly.

Now, we note that the $\phi$ interval spanned by the obstruction depends both on $r_o$ and on the $\mu$-value of a slice, i.e. $|\phi| < \phi_{max}(r_o, \mu)$. The limiting value $\phi_{max}(r_o, \mu)$ is found by setting $\rho_-(\mu, \phi_{max}) = \rho_+(\mu, \phi_{max})$, which results in:

$$0 = \sqrt{(r_o^2 \cos^2 \phi_{max} - r_p^2) \cosh^2 \mu \sinh^2 \mu + (r_s^2 - r_o^2 \sin^2 \phi_{max}) \cosh^2 \mu}$$

$$0 = r_o^2 \cos^2 \phi_{max} \sinh^2 \mu - r_p^2 \sinh^2 \mu + r_s^2 - r_o^2 \sin^2 \phi_{max}$$

$$0 = (r_o^2 - r_p^2) \sinh^2 \mu + r_s^2 - r_o^2 \sin^2 \phi_{max} \cosh^2 \mu$$

$$\sin^2 \phi_{max} = \left(1 - \frac{r_p^2}{r_o^2}\right) \tanh^2 \mu + \frac{r_s^2}{r_o^2} \text{sech}^2 \mu$$

$$\Rightarrow \phi_{max} = \sin^{-1}\left(\sqrt{\left(1 - \frac{r_p^2 + r_s^2}{r_o^2}\right) \tanh^2 \mu + \frac{r_s^2}{r_o^2}}\right) \tag{S59}$$

Lastly, we find the maximal $\mu$-value containing the sphere, $\mu_{max}$, in order to delimit the obstructed and unobstructed $\mu$-integrals. $\mu_{max}$ is found by setting both $\rho_- = \rho_+$ and $\phi = 0$:

$$\mu_{max} = \sinh^{-1}\left(\sqrt{\frac{r_s^2}{r_p^2 - r_o^2}}\right) \tag{S60}$$

For $r_o > r_s$, the resistance of the obstructed oblate spheroidal segment is therefore:

$$R'_{sph}(r_o > r_s) = \frac{2}{\sigma r_p} \int_0^{\mu_{max}} \frac{\text{sech}\,\mu\, d\mu}{\int_{\phi_{max}}^{2\pi - \phi_{max}} [\sin v]_0^{\frac{\pi}{2}} d\phi + \int_{-\phi_{max}}^{\phi_{max}} \left([\sin v]_0^{v_+} + [\sin v]_{v_-}^{\frac{\pi}{2}}\right) d\phi}$$

$$= \frac{2}{\sigma r_p} \int_0^{\sinh^{-1}\left(\frac{r_s}{\sqrt{r_p^2 - r_o^2}}\right)} \frac{\text{sech}\,\mu\, d\mu}{2\pi + \int_{-\phi_{max}}^{\phi_{max}} \left(\sqrt{1 - \frac{\rho_+^2}{r_p^2}\text{sech}^2\,\mu} - \sqrt{1 - \frac{\rho_-^2}{r_p^2}\text{sech}^2\,\mu}\right) d\phi} \tag{S61}$$

And for $r_o < r_s$:

$$R'_{sph}(r_o < r_s) = \frac{2}{\sigma r_p} \int_0^{\sinh^{-1}\left(\frac{\sqrt{r_s^2 - r_o^2}}{r_p}\right)} \frac{\text{sech}\,\mu\, d\mu}{\int_0^{2\pi} \sqrt{1 - \frac{\rho_+^2}{r_p^2}\text{sech}^2\,\mu}\, d\phi}$$

$$+ \frac{2}{\sigma r_p} \int_{\sinh^{-1}\left(\frac{\sqrt{r_s^2 - r_o^2}}{r_p}\right)}^{\sinh^{-1}\left(\frac{r_s}{\sqrt{r_p^2 - r_o^2}}\right)} \frac{\text{sech}\,\mu\, d\mu}{2\pi + \int_{-\phi_{max}}^{\phi_{max}} \left(\sqrt{1 - \frac{\rho_+^2}{r_p^2}\text{sech}^2\,\mu} - \sqrt{1 - \frac{\rho_-^2}{r_p^2}\text{sech}^2\,\mu}\right) d\phi} \tag{S62}$$

The conductance of the entire system considers the contributions from the unobstructed segments:

$$\frac{G_b^{sph}}{G_o}(r_o, r_s) = \left[1 - \frac{2}{\pi}\tan^{-1}\left(\frac{r_s}{\sqrt{r_p^2 - r_o^2}}\right) + 2\sigma r_p R'_{sph}(r_o, r_s)\right]^{-1} \tag{S63}$$

The dependence of Equation S63 on $r_o/r_p$ is plotted in Figure S10a for different sphere radii alongside the corresponding values obtained from finite-element simulations. Figure S10b further plots the predicted and simulated excess conductance at $r_o = 0$ compared to nonzero $r_o$ values. Interestingly, although the absolute values predicted from equation S63 are in worse agreement with the simulations than those of cylinders (Fig. S8a), the relative $r_o$-dependence compared to a centred obstruction is better captured for spheres than for cylinders, as evidenced by the closer spacing between modelled and simulated values in Figure S10b (compare to Fig. S8b).

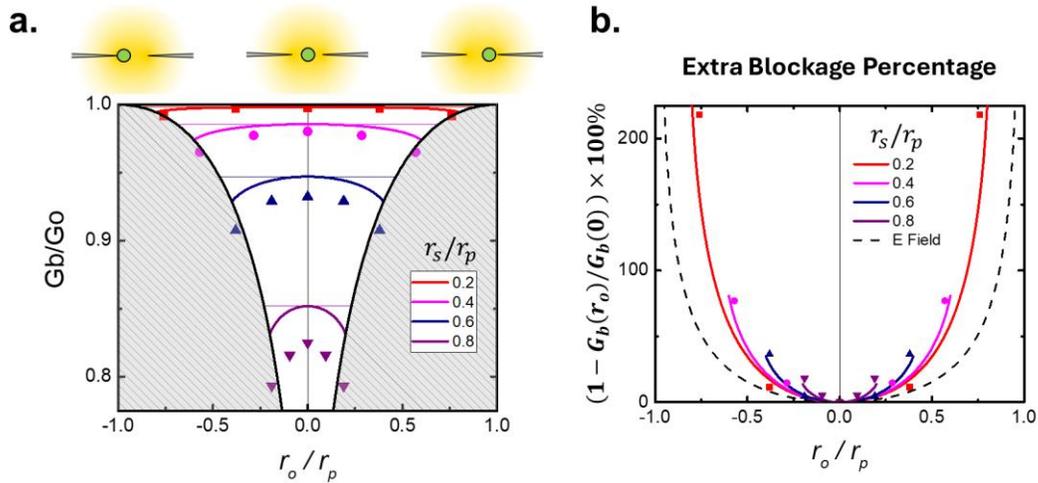

**Figure S11.** Off-axis effects for vertically centred spherical obstructions. **a)** Dependence of normalized blocked-state conductance $G_b^{sph}/G_o$ on radial position $r_o$ of the sphere inside the pore. **b)** Extra conductance induced by centred sphere position $r_o = 0$, compared to off-axis $r_o \neq 0$. All individual points are values calculated from finite element simulations, whereas continuous curves are from Equations S61-S63.

## S9. Complex Obstruction – More Examples

As discussed in the main text (Figure 4), the presented modeling framework allows for predicting the shape of signals from real-life complex molecules by first modeling them as a sequence of sub-units with simple geometries. Here, we show a few examples to demonstrate how this can be applied to various types of molecules.

### Side Chains on Scaffold

We first show how to address modeling the nanopore signals of molecules used in scaffold-assisted sensing, wherein molecules of interest such as RNA, proteins, or biomarkers are bound along an extended double-stranded DNA backbone and produce an additional transient blockage during their passage through the pore. Here, we model the passage of a cylindrical backbone $r_c = 0.4r_p$, with 7 equidistant side chains of length $L_{side} = 10r_p$ separated by $L_{space} = 100r_p$. Figure S11 shows three different methods to model sidechains: i) as cylinders with radii $r_{side} = 0.6r_p$, ii) as wedged cylinders with $r_1 = 0.4r_p$, $r_2 = 0.83r_p$, $\phi_{side} = 3\pi/4$, and iii) as spheroids with $a = L_{side}$ and

$b = 0.6r_p$. These specific values were chosen so that the maximal cross-sectional area of each type of side chain would be kept constant.

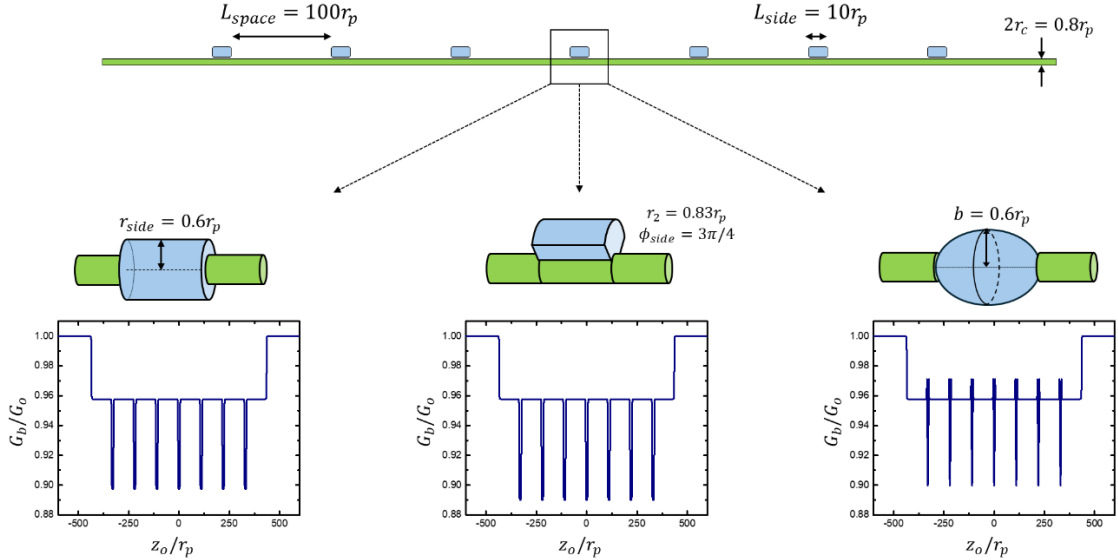

**Figure S12.** Different ways of modeling seven sidechains along a cylindrical scaffold: **a)** cylindrical, **b)** wedged cylinder, and **c)** spheroidal obstructions. In each case, the modeled conductance blockages $G_b$ are plotted as a function of the distance $z_o$ of the scaffold centre from the 2D pore opening (radius $r_p$).[1]

**Spatial Resolution Limit for Molecular Design**

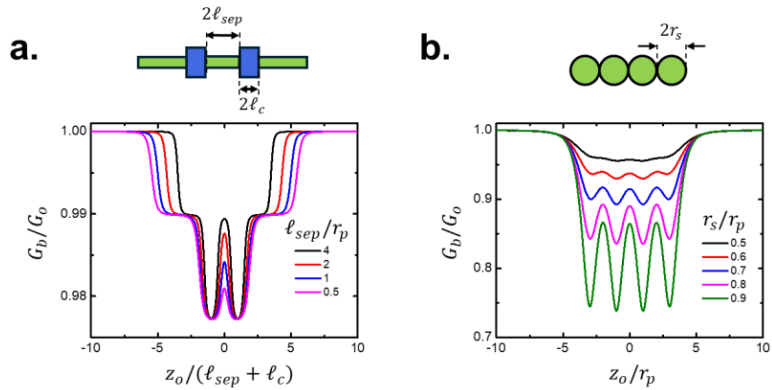

**Figure S13.** Inherent resolution limit demonstrated with **a)** short barcode-like cylinders in series ($r_c^{wide} = 0.3r_p$, $r_c^{narrow} = 0.2r_p$, $\ell_c = 2r_p$), and **b)** spheres in series.

We now comment on the intrinsic $z$-resolution limit of the components of a complex obstruction, imposed by the non-zero sensing volume of a 2D pore. Figure S13a shows the conductance trace of a barcode-like obstruction made up of five cylinders in series, with two wider but shorter ($r_c^{wide} = 0.3r_p$, $L_c = 2\ell_c = 4r_p$) sub-units separated by $L_{sep} = 2\ell_{sep}$ along a backbone with $r_c^{narrow} = 0.2r_p$. By changing the value of $\ell_{sep}$, Figure S13 shows the that the larger diameter features are harder to resolve within a $z$-mapping when they are separated by less than $2r_p$, again implying that the sensing volume of a 2D pore of radius $r_p$ has dimension $2r_p$ in the $z$ direction. Similarly, Figure S13b shows the conductance $z$-mappings expected from a series of four identical spheres with radii ranging between $r_s = 0.5r_p$ and $r_s = 0.9r_p$. Although spheres that take up most of the sensing volume ($r_s \approx r_p$) can be identified individually, strings of smaller spheres result in traces without obviously identifiable sub-unit characteristics. The observations from Figure S13 provide useful insights with regards to designing molecules with information contained in their substructure to be sensed with pores.

**Insects Wearing Hats**

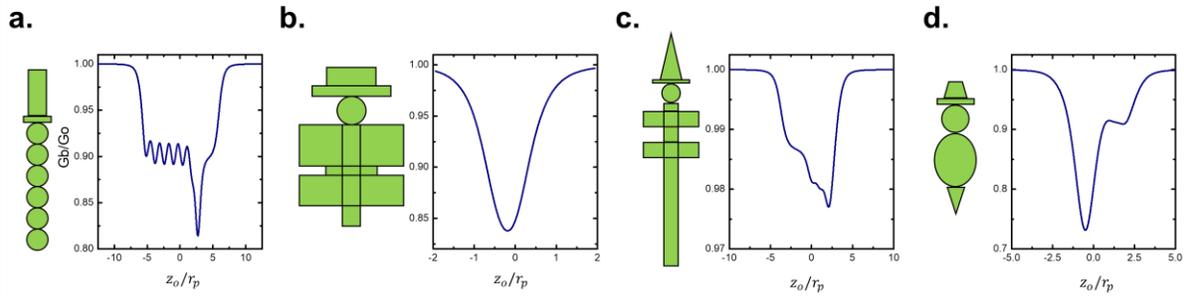

**Figure S14.** Extra demonstration of blockade predictions for complex obstructions by modeling simple sub-units in series. Conductance expected during the passage through a 2D pore of **a)** Caterpillar wearing a top hat; **b)** Butterfly wearing a boater hat; **c)** Dragonfly wearing a witch hat; **d)** Bee wearing a fedora hat.

**S10. Finite-Length Pore Equations**

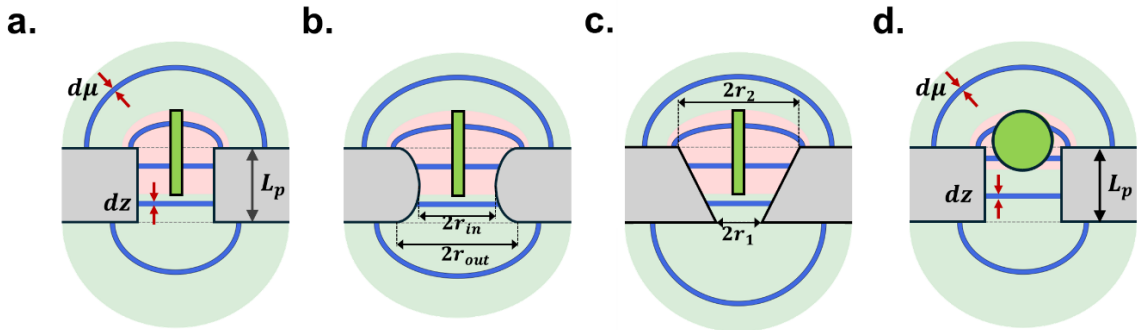

**Figure S15.** Method for modeling conductance of finite-length pores: oblate spheroidal slices outside the pore and circular slices inside the pore. **a)** Cylindrical obstruction in a cylindrical pore. **b)** Cylindrical obstruction in a hyperboloidal pore. **c)** Cylindrical obstruction in a conical pore. **d)** Spherical obstruction in a cylindrical pore.

As discussed in the main text, the oblate spheroidal resistive slice method can also be applied to estimate the access resistance of finite-length pores. The total conductance of a pore of radius $r_p$ and length $L_p = 2\ell_p$ is thus calculated by considering the contributions from the two access regions and channel region in series:

$$G_b = \left[R_{acc}^{bot} + R_{pore} + R_{acc}^{top}\right]^{-1}$$
$$= \left[\int_{-\infty}^{0} dR_{acc}^{bot}(\mu) + \int_{-\ell_p}^{\ell_p} dR_{pore}(z) + \int_{0}^{\infty} dR_{acc}^{top}(\mu)\right]^{-1} \quad \text{(S64)}$$

For the simple scenario of a cylindrical obstruction of radius $r_c$ and length $L_c = 2\ell_c$ at a distance of $z_o$ from a pore with a $z$-dependent radial profile of $r_p(z)$, and of length $L_p = 2\ell_p$, the terms of equation S64 can be further segmented into obstructed and unobstructed regions. For this reason, the resistance calculation needs to be broken down into five scenarios, depending on whether the cylindrical obstruction is: i) completely in the top access region $(z_o > \ell_p + \ell_c)$, ii) partially in the pore and the top access region $(|\ell_p - \ell_c| < z_o < \ell_p + \ell_c)$, iii) entirely inside the pore $(\ell_c - \ell_p < z_o < \ell_p - \ell_c$ for $\ell_c < \ell_p)$ or partially inside the top access region, pore, and bottom access region $(\ell_p - \ell_c < z_o < \ell_c - \ell_p$ for $\ell_c > \ell_p)$, iv) partially in the pore and the bottom access region $(-\ell_p - \ell_c < z_o < -|\ell_p - \ell_c|)$, or v) entirely in the bottom access region $(z_o < -\ell_p - \ell_c)$. Note that in the case of $z$-symmetric pore shapes, the number of scenarios to consider drops to three. Let us now consider the solutions for cylindrical obstructions in channels of various rotationally symmetric shapes.

**Finite Cylinder in Cylindrical Pore**

We first consider a cylindrical pore with a radial profile trivially parametrized as $r_p(z) = r_p$. Note that the access region contributions are identical to Equation 10 from the main text, only with a translation in $z_o$ of $\pm \ell_p$ due to the presence of the membrane. The infinitesimal resistance of thin obstructed slices inside and outside the pore are thus calculated as:

$$dR'_{pore}(z) = \frac{dz}{\sigma\pi(r_p^2 - r_c^2)} \quad \text{(S65)}$$

$$dR'_{acc}(\mu) = \frac{\text{sech}\,\mu\, d\mu}{2\pi\sigma r_p \sqrt{1 - \frac{r_c^2}{r_p^2}\text{sech}^2\,\mu}} \quad \text{(S66)}$$

Open-Pore

$$R_o = \frac{2\ell_p}{\sigma\pi r_p^2} + \frac{1}{2\sigma r_p} \quad \text{(S67)}$$

Cylinder Outside the pore $(\ell_p + \ell_c < |z_o|)$:

$$R_b(z) = \frac{2\ell_p}{\sigma\pi r_p^2} + \frac{1}{2\sigma r_p}\left[1 + \frac{1}{\pi}\tan^{-1}\left(\frac{|z_o| - \ell_p - \ell_c}{r_p}\right) - \frac{1}{\pi}\tan^{-1}\left(\frac{|z_o| - \ell_p + \ell_c}{r_p}\right)\right.$$

$$+\frac{1}{\pi}F\left(\frac{\pi}{2}-\tan^{-1}\left(\frac{|z_o|-\ell_p-\ell_c}{r_p}\right),\frac{r_c}{r_p}\right)-\frac{1}{\pi}F\left(\frac{\pi}{2}-\tan^{-1}\left(\frac{|z_o|-\ell_p+\ell_c}{r_p}\right),\frac{r_c}{r_p}\right)\right] \quad \text{(S68)}$$

Cylinder Partially inside the pore $\left(|\ell_p-\ell_c|<|z_o|<\ell_p+\ell_c\right)$:

$$R_b(z_o) = \frac{\ell_p-|z_o|+\ell_c}{\sigma\pi(r_p^2-r_c^2)} + \frac{\ell_p+|z_o|-\ell_c}{\sigma\pi r_p^2} + \frac{1}{2\sigma r_p}\left[1-\frac{1}{\pi}\tan^{-1}\left(\frac{|z_o|+\ell_c-\ell_p}{r_p}\right)\right.$$
$$\left.+\frac{1}{\pi}F\left(\frac{\pi}{2},\frac{r_c}{r_p}\right)-\frac{1}{\pi}F\left(\frac{\pi}{2}-\tan^{-1}\left(\frac{|z_o|+\ell_c-\ell_p}{r_p}\right),\frac{r_c}{r_p}\right)\right] \quad \text{(S69)}$$

Long cylinder $(\ell_c>\ell_p)$ inside the pore $\left(0<|z_o|<\ell_c-\ell_p\right)$:

$$R_b(z_o) = \frac{2\ell_p}{\sigma\pi(r_p^2-r_c^2)} + \frac{1}{2\sigma r_p}\left[1+\frac{1}{\pi}\tan^{-1}\left(\frac{|z_o|-\ell_c+\ell_p}{r_p}\right)-\frac{1}{\pi}\tan^{-1}\left(\frac{|z_o|+\ell_c-\ell_p}{r_p}\right)\right.$$
$$\left.+\frac{1}{\pi}F\left(\frac{\pi}{2}-\tan^{-1}\left(\frac{|z_o|-\ell_c+\ell_p}{r_p}\right),\frac{r_c}{r_p}\right)-\frac{1}{\pi}F\left(\frac{\pi}{2}-\tan^{-1}\left(\frac{|z_o|+\ell_c-\ell_p}{r_p}\right),\frac{r_c}{r_p}\right)\right] \quad \text{(S70)}$$

Short cylinder $(\ell_c<\ell_p)$ inside the pore $\left(0<|z_o|<\ell_p-\ell_c\right)$:

$$R_b(z_o) = \frac{2\ell_c}{\sigma\pi(r_p^2-r_c^2)} + \frac{2(\ell_p-\ell_c)}{\sigma\pi r_p^2} + \frac{1}{2\sigma r_p} \quad \text{(S71)}$$

Interactive Desmos plot (for $z_o=0$): https://www.desmos.com/calculator/vocbsxz6yn

Very long cylinder $(\ell_c \gg \ell_p)$ centered inside pore:

As it is relevant for semiflexible linear polymers with a persistence length longer than that of the pore, such as for double stranded DNA in a nanoscale pore, we also note the solution for very long cylindrical obstructions ($\ell_c \gg \ell_p$):

$$R_b = \frac{2\ell_p}{\sigma\pi(r_p^2-r_c^2)} + \frac{\frac{2}{\pi}K\left(\frac{r_c}{r_p}\right)}{2\sigma r_p} \quad \text{(S72)}$$

Here $K(k)=F\left(\frac{\pi}{2},k\right)$ is the complete elliptic integral of the first kind. Equation S72 and its accuracy in various conditions has been evaluated and discussed extensively in a recent publication.[1]

**Finite Cylinder in Hyperboloidal Pore**

We now consider a cylindrical obstruction of length $L_c=2\ell_c$ and radius $r_c$ passing through a hyperboloidal pore of length $L_p=2\ell_p$ characterized by its inner and outer radii $r_{in}$ and $r_{out}$ (See Figure S15) described by the following parametrization:

$$r_p^2(z) = r_{in}^2 + \frac{r_{out}^2-r_{in}^2}{\ell_p^2}z^2 \quad \text{(S73)}$$

The infinitesimal resistance of thin obstructed slices inside and outside the pore are calculated as:

$$dR'_{pore}(z) = \frac{dz}{\sigma\pi\left(r_{in}^2 - r_c^2 + \frac{r_{out}^2 - r_{in}^2}{\ell_p^2}z^2\right)} \tag{S74}$$

$$dR'_{acc}(\mu) = \frac{\text{sech}\,\mu\,d\mu}{2\pi\sigma r_{out}\sqrt{1 + \frac{r_c^2}{r_{out}^2}\text{sech}^2\,\mu}} \tag{S75}$$

Open-Pore

$$R_o = \frac{2\ell_p}{\sigma\pi r_{in}\sqrt{r_{out}^2 - r_{in}^2}}\tan^{-1}\sqrt{\frac{r_{out}^2}{r_{in}^2} - 1} + \frac{1}{2\sigma r_{out}} \tag{S76}$$

Cylinder Outside the pore $(\ell_p + \ell_c < |z_o|)$:

$$R_b(z_o) = \frac{2\ell_p}{\sigma\pi r_{in}\sqrt{r_{out}^2 - r_{in}^2}}\tan^{-1}\sqrt{\frac{r_{out}^2}{r_{in}^2} - 1}$$
$$+ \frac{1}{2\sigma r_{out}}\left[1 + \frac{1}{\pi}\tan^{-1}\left(\frac{|z_o| - \ell_c - \ell_p}{r_{out}}\right) - \frac{1}{\pi}\tan^{-1}\left(\frac{|z_o| + \ell_c - \ell_p}{r_{out}}\right)\right.$$
$$\left. + \frac{1}{\pi}F\left(\frac{\pi}{2} - \tan^{-1}\left(\frac{|z_o| - \ell_p - \ell_c}{r_{out}}\right), \frac{r_c}{r_{out}}\right) - \frac{1}{\pi}F\left(\frac{\pi}{2} - \tan^{-1}\left(\frac{|z_o| - \ell_p + \ell_c}{r_{out}}\right), \frac{r_c}{r_{out}}\right)\right] \tag{S77}$$

Cylinder Partially inside the pore $(|\ell_p - \ell_c| < |z_o| < \ell_p + \ell_c)$:

$$R_b(z_o) = \frac{\ell_p}{\sigma\pi\sqrt{r_{out}^2 - r_{in}^2}}\left[\frac{1}{r_{in}}\tan^{-1}\left(\sqrt{\frac{r_{out}^2}{r_{in}^2} - 1}\left(\frac{|z_o|z_o - \ell_c}{\ell_p}\right)\right) + \frac{1}{r_{in}}\tan^{-1}\left(\sqrt{\frac{r_{out}^2}{r_{in}^2} - 1}\right)\right.$$
$$\left. + \frac{1}{\sqrt{r_{in}^2 - r_c^2}}\tan^{-1}\left(\sqrt{\frac{r_{out}^2 - r_{in}^2}{r_{in}^2 - r_c^2}}\right) - \frac{1}{\sqrt{r_{in}^2 - r_c^2}}\tan^{-1}\left(\sqrt{\frac{r_{out}^2 - r_{in}^2}{r_{in}^2 - r_c^2}}\left(\frac{|z_o| - \ell_c}{\ell_p}\right)\right)\right]$$
$$+ \frac{1}{2\sigma r_{out}}\left[1 - \frac{1}{\pi}\tan^{-1}\left(\frac{|z_o| + \ell_c - \ell_p}{r_{out}}\right) + \frac{1}{\pi}F\left(\frac{\pi}{2}, \frac{r_c}{r_{out}}\right)\right.$$
$$\left. - \frac{1}{\pi}F\left(\frac{\pi}{2} - \tan^{-1}\left(\frac{|z_o| + \ell_c - \ell_p}{r_{out}}\right), \frac{r_c}{r_{out}}\right)\right] \tag{S78}$$

Long cylinder $(\ell_c > \ell_p)$ inside the pore $(|z_o| < \ell_c - \ell_p)$:

$$R_b(z_o) = \frac{2\ell_p}{\sigma\pi\sqrt{(r_{out}^2 - r_{in}^2)(r_{in}^2 - r_c^2)}}\tan^{-1}\left(\sqrt{\frac{r_{out}^2 - r_{in}^2}{r_{in}^2 - r_c^2}}\right)$$
$$+ \frac{1}{2\sigma r_{out}}\left[1 + \frac{1}{\pi}\tan^{-1}\left(\frac{|z_o| - \ell_c + \ell_p}{r_{out}}\right) - \frac{1}{\pi}\tan^{-1}\left(\frac{|z_o| + \ell_c - \ell_p}{r_{out}}\right)\right.$$
$$\left. + \frac{1}{\pi}F\left(\frac{\pi}{2} - \tan^{-1}\left(\frac{|z_o| - \ell_c + \ell_p}{r_{out}}\right), \frac{r_c}{r_{out}}\right) - \frac{1}{\pi}F\left(\frac{\pi}{2} - \tan^{-1}\left(\frac{|z_o| + \ell_c - \ell_p}{r_{out}}\right), \frac{r_c}{r_{out}}\right)\right] \tag{S79}$$

Short cylinder $(\ell_c < \ell_p)$ inside the pore $(|z_o| < \ell_p - \ell_c)$:

$$R_b(z_o) = \frac{1}{2\sigma r_{out}} + \frac{L}{\sigma\pi\sqrt{r_{out}^2 - r_{in}^2}}\left[\frac{2}{r_{in}}\tan^{-1}\left(\sqrt{\frac{r_{out}^2 - r_{in}^2}{r_{in}^2}}\right) + \frac{1}{r_{in}}\tan^{-1}\left(\sqrt{\frac{r_{out}^2}{r_{in}^2} - 1}\left(\frac{|z_o| - \ell_c}{\ell_p}\right)\right)\right.$$

$$\left. - \frac{1}{r_{in}}\tan^{-1}\left(\sqrt{\frac{r_{out}^2}{r_{in}^2} - 1}\left(\frac{|z_o| + \ell_c}{\ell_p}\right)\right) - \frac{1}{\sqrt{r_{in}^2 - r_c^2}}\tan^{-1}\left(\sqrt{\frac{r_{out}^2 - r_{in}^2}{r_{in}^2 - r_c^2}}\left(\frac{|z_o| - \ell_c}{\ell_p}\right)\right)\right.$$

$$\left. + \frac{1}{\sqrt{r_{in}^2 - r_c^2}}\tan^{-1}\left(\sqrt{\frac{r_{out}^2 - r_{in}^2}{r_{in}^2 - r_c^2}}\left(\frac{|z_o| + \ell_c}{\ell_p}\right)\right)\right] \quad (S80)$$

Interactive Desmos plot (for $z_o = 0$): https://www.desmos.com/calculator/lymjxdn9kj

Very long cylinder $(\ell_c \gg \ell_p)$ centered inside pore:

$$R_b(z_o = 0) = \frac{2\ell_p}{\sigma\pi\sqrt{(r_{out}^2 - r_{in}^2)(r_{in}^2 - r_c^2)}}\tan^{-1}\left(\sqrt{\frac{r_{out}^2 - r_{in}^2}{r_{in}^2 - r_c^2}}\right) + \frac{\frac{2}{\pi}K\left(\frac{r_c}{r_{out}}\right)}{2\sigma r_{out}} \quad (S81)$$

**Finite Cylinder in Conical Pore**

We now consider a cylindrical obstruction of length $L_c = 2\ell_c$ and radius $r_c$ passing through a conical pore of length $L_p = 2\ell_p$ characterized by its lower and upper radii $r_1$ and $r_2$ (See Figure S15) described with the following parametrization:

$$r_p(z) = \frac{r_2 - r_1}{2\ell_p}z + \frac{r_1 + r_2}{2} \quad (S82)$$

The infinitesimal resistance of thin obstructed slices inside and outside the pore are calculated as:

$$dR'_{pore}(z) = \frac{dz}{\sigma\pi\left(\left(\frac{r_2 - r_1}{2\ell_p}z + \frac{r_1 + r_2}{2}\right)^2 - r_c^2\right)} \quad (S83)$$

$$dR'^{top}_{acc}(\mu) = \frac{\text{sech}\,\mu\, d\mu}{2\pi\sigma r_2\sqrt{1 + \frac{r_c^2}{r_2^2}\text{sech}^2\mu}} \quad (S84)$$

$$dR'^{bot}_{acc}(\mu) = \frac{\text{sech}\,\mu\, d\mu}{2\pi\sigma r_1\sqrt{1 + \frac{r_c^2}{r_1^2}\text{sech}^2\mu}} \quad (S85)$$

Open-Pore

$$R_o = \frac{1}{4\sigma r_1} + \frac{1}{4\sigma r_2} + \frac{2\ell_p}{\pi \sigma r_1 r_2} \tag{S86}$$

Cylinder above the pore $(\ell_p + \ell_c < z_o)$:

$$\begin{aligned} R_b(z_o) = & \frac{1}{4\sigma r_1} + \frac{2\ell_p}{\pi \sigma r_1 r_2} \\ & + \frac{1}{4\sigma r_2}\left[1 + \frac{2}{\pi}\tan^{-1}\left(\frac{z_o - \ell_c - \ell_p}{r_2}\right) - \frac{2}{\pi}\tan^{-1}\left(\frac{z_o + \ell_c - \ell_p}{r_2}\right)\right. \\ & \left. + \frac{2}{\pi} F\left(\frac{\pi}{2} - \tan^{-1}\left(\frac{z_o - \ell_c - \ell_p}{r_2}\right), \frac{r_c}{r_2}\right) - \frac{2}{\pi} F\left(\frac{\pi}{2} - \tan^{-1}\left(\frac{z_o + \ell_c - \ell_p}{r_2}\right), \frac{r_c}{r_2}\right)\right] \end{aligned} \tag{S87}$$

Cylinder above and partially inside the pore $(|\ell_c - \ell_p| < z_o < \ell_p + \ell_c)$:

$$\begin{aligned} R_b(z_o) = & \frac{1}{4\sigma r_1} + \frac{2\ell_p}{\pi\sigma(r_2 - r_1)}\left[\frac{1}{r_1} - \frac{2}{\frac{r_2 - r_1}{\ell_p}(z_o - \ell_c) + r_1 + r_2}\right. \\ & \left. + \frac{1}{2r_c}\ln\left(\frac{(r_2 - r_c)\left(\frac{r_2 - r_1}{\ell_p}(z_o - \ell_c) + r_1 + r_2 + 2r_c\right)}{(r_2 + r_c)\left(\frac{r_2 - r_1}{\ell_p}(z_o - \ell_c) + r_1 + r_2 - 2r_c\right)}\right)\right] \\ & + \frac{1}{4\sigma r_2}\left[1 - \frac{2}{\pi}\tan^{-1}\left(\frac{z_o + \ell_c - \ell_p}{r_2}\right) + \frac{2}{\pi}F\left(\frac{\pi}{2}, \frac{r_c}{r_2}\right)\right. \\ & \left. - \frac{2}{\pi}F\left(\frac{\pi}{2} - \tan^{-1}\left(\frac{z_o + \ell_c - \ell_p}{r_2}\right), \frac{r_c}{r_2}\right)\right] \end{aligned} \tag{S88}$$

Long cylinder $(\ell_c > \ell_p)$ inside the pore $(|z_o| < \ell_c - \ell_p)$:

$$\begin{aligned} R_b(z_o) = & \frac{1}{4\sigma r_1}\left(1 + \frac{2}{\pi}\tan^{-1}\left(\frac{z_o - \ell_c + \ell_p}{r_1}\right) + \frac{2}{\pi}F\left(\frac{\pi}{2} - \tan^{-1}\left(\frac{z_o - \ell_c + \ell_p}{r_1}\right), \frac{r_c}{r_1}\right) - \frac{2}{\pi}F\left(\frac{\pi}{2}, \frac{r_c}{r_1}\right)\right) \\ & + \frac{1}{4\sigma r_2}\left(1 - \frac{2}{\pi}\tan^{-1}\left(\frac{z_o + \ell_c - \ell_p}{r_2}\right) - \frac{2}{\pi}F\left(\frac{\pi}{2} - \tan^{-1}\left(\frac{z_o + \ell_c - \ell_p}{r_2}\right), \frac{r_c}{r_2}\right) + \frac{2}{\pi}F\left(\frac{\pi}{2}, \frac{r_c}{r_2}\right)\right) \\ & + \frac{\ell_p}{\pi\sigma r_c(r_2 - r_1)}\ln\left(\frac{(r_2 - r_c)(r_1 + r_c)}{(r_2 + r_c)(r_1 - r_c)}\right) \end{aligned} \tag{S89}$$

Short cylinder $(\ell_c < \ell_p)$ inside the pore $(|z_o| < \ell_p - \ell_c)$:

$$\begin{aligned} R_b(z_o) = & \frac{2\ell_p}{\pi\sigma(r_2 - r_1)}\left[\frac{1}{r_1} - \frac{1}{r_2} + \frac{2}{\frac{r_2 - r_1}{\ell_p}(z_o + \ell_c) + r_1 + r_2} - \frac{2}{\frac{r_2 - r_1}{\ell_p}(z_o - \ell_c) + r_1 + r_2}\right. \\ & \left. + \frac{1}{2r_c}\ln\left(\frac{\left(\frac{r_2 - r_1}{\ell_p}(z_o + \ell_c) + r_1 + r_2 - 2r_c\right)\left(\frac{r_2 - r_1}{\ell_p}(z_o - \ell_c) + r_1 + r_2 + 2r_c\right)}{\left(\frac{r_2 - r_1}{\ell_p}(z_o + \ell_c) + r_1 + r_2 + 2r_c\right)\left(\frac{r_2 - r_1}{\ell_p}(z_o - \ell_c) + r_1 + r_2 - 2r_c\right)}\right)\right] \\ & + \frac{1}{4\sigma r_1} + \frac{1}{4\sigma r_2} \end{aligned} \tag{S90}$$

Cylinder under and partially inside the pore $\left(-|\ell_c - \ell_p| > z_o > -(\ell_p + \ell_c)\right)$:

$$R_b(z_o) = \frac{1}{4\sigma r_2} + \frac{2\ell_p}{\pi\sigma(r_2 - r_1)} \left[ \frac{2}{\frac{r_2 - r_1}{\ell_p}(z_o + \ell_c) + r_1 + r_2} - \frac{1}{r_2} \right.$$

$$+ \frac{1}{2r_c} \ln \left( \frac{\left(\frac{r_2 - r_1}{\ell_p}(z_o + \ell_c) + r_1 + r_2 - 2r_c\right)(r_1 + r_c)}{\left(\frac{r_2 - r_1}{\ell_p}(z_o + \ell_c) + r_1 + r_2 + 2r_c\right)(r_1 - r_c)} \right) \Bigg]$$

$$+ \frac{1}{4\sigma r_1} \left[ 1 + \frac{2}{\pi} \tan^{-1}\left(\frac{z_o - \ell_c + \ell_p}{r_1}\right) - \frac{2}{\pi} F\left(\frac{\pi}{2}, \frac{r_c}{r_1}\right) \right.$$

$$\left. + \frac{2}{\pi} F\left(\frac{\pi}{2} - \tan^{-1}\left(\frac{z_o - \ell_c + \ell_p}{r_1}\right), \frac{r_c}{r_1}\right) \right] \qquad \text{(S91)}$$

Cylinder under the pore $\left(z_o < -(\ell_p + \ell_c)\right)$:

$$R_b(z_o) = \frac{2\ell_p}{\pi\sigma r_1 r_2} + \frac{1}{4\sigma r_2}$$

$$+ \frac{1}{4\sigma r_1} \left[ 1 + \frac{2}{\pi} \tan^{-1}\left(\frac{z_o - \ell_c + \ell_p}{r_1}\right) - \frac{2}{\pi} \tan^{-1}\left(\frac{z_o + \ell_c + \ell_p}{r_1}\right) \right.$$

$$\left. + \frac{2}{\pi} F\left(\frac{\pi}{2} - \tan^{-1}\left(\frac{z_o - \ell_c + \ell_p}{r_1}\right), \frac{r_c}{r_1}\right) - \frac{2}{\pi} F\left(\frac{\pi}{2} - \tan^{-1}\left(\frac{z_o + \ell_c + \ell_p}{r_1}\right), \frac{r_c}{r_1}\right) \right] \qquad \text{(S92)}$$

Interactive Desmos plot (for $z_o = 0$): https://www.desmos.com/calculator/xewxe1svij

Very long cylinder $(\ell_c \gg \ell_p)$ centered inside pore:

$$R_b = \frac{K\left(\frac{r_c}{r_1}\right)}{2\pi\sigma r_1} + \frac{K\left(\frac{r_c}{r_2}\right)}{2\pi\sigma r_2} + \frac{\ell_p}{\pi\sigma r_c(r_2 - r_1)} \ln\left(\frac{(r_2 - r_c)(r_1 + r_c)}{(r_2 + r_c)(r_1 - r_c)}\right) \qquad \text{(S93)}$$

**Sphere in Cylindrical Pore**

As discussed in Section S4, the infinitesimal resistance of a constant-$\mu$ oblate spheroidal slice in the presence of a spherical obstruction (radius $r_s$) at a distance of $z_o$ from a pore in a 2D membrane is (see Eq. S32):

$$dR'_{2D}(z_o) = \frac{1}{2\pi\sigma r_p} \frac{\text{sech}^2 \mu \, d\mu}{\sqrt{1 - \frac{r_s^2 + z_o^2}{r_p^2} \text{sech}^2 \mu + 2\frac{z_o^2}{r_p^2} - 2\frac{z_o}{r_p}\tanh\mu \sqrt{1 + \frac{z_o^2}{r_p^2} - \frac{r_s^2}{r_p^2}\text{sech}^2 \mu}}} \qquad \text{(S94)}$$

When treating oblate spheroidal slices in the access regions of finite-length pores, Equation S93 can be adapted through a corresponding $z$-translation:

$$dR'_{acc}(z_o) = dR'_{2D}(|z_o| - \ell_p) \tag{S95}$$

The surface of the spherical obstruction is parametrized with $r_o^2 = r_s^2 - (z - z_o)^2$, and thus the resistance of an obstructed circular slice inside the pore is evaluated as:

$$dR'_{pore}(z_o) = \frac{dz}{\sigma\pi(r_p^2 - r_s^2 + (z - z_o)^2)} = \frac{1}{\sigma\pi(r_p^2 - r_s^2)} \frac{dz}{1 + \frac{(z - z_o)^2}{r_p^2 - r_s^2}} \tag{S96}$$

<u>Sphere outside pore</u> $(\ell_p + r_s < |z_o|)$:

$$R_b(z_o) = \frac{2\ell_p}{\sigma\pi r_p^2} + \int_{\sinh^{-1}\left(\frac{|z_o|-r_s-\ell_p}{r_p}\right)}^{\sinh^{-1}\left(\frac{|z_o|+r_s-\ell_p}{r_p}\right)} dR'_{2D}(|z_o| - \ell_p)$$
$$+ \frac{1}{2\sigma r_p}\left[1 - \frac{1}{\pi}\tan^{-1}\left(\frac{|z_o|+r_s-\ell_p}{r_p}\right) + \frac{1}{\pi}\tan^{-1}\left(\frac{|z_o|-r_s-\ell_p}{r_p}\right)\right] \tag{S97}$$

<u>Sphere partially inside pore</u> $(|\ell_p - r_s| < |z_o| < \ell_p + r_s)$:

$$R_b(z_o) = \frac{\ell_p + |z_o| - r_s}{\sigma\pi r_p^2} + \frac{1}{2\sigma r_p}\left(1 - \frac{1}{\pi}\tan^{-1}\left(\frac{|z_o|+r_s-\ell_p}{r_p}\right)\right)$$
$$+ \frac{1}{\sigma\pi\sqrt{r_p^2 - r_s^2}}\left(\tan^{-1}\left(\frac{\ell_p - |z_o|}{\sqrt{r_p^2 - r_s^2}}\right) + \sin^{-1}\left(\frac{r_s}{r_p}\right)\right)$$
$$+ \int_0^{\sinh^{-1}\left(\frac{|z_o|+r_s-\ell_p}{r_p}\right)} dR'_{2D}(|z_o| - \ell_p) \tag{S98}$$

<u>Large sphere</u> $(r_s > \ell_p)$ <u>inside the pore</u> $(|z_o| < r_s - \ell_p)$:

$$R_b(z_o) = \frac{1}{2\sigma r_p}\left(1 + \frac{1}{\pi}\tan^{-1}\left(\frac{z_o - r_s + \ell_p}{r_p}\right) + \frac{1}{\pi}\tan^{-1}\left(\frac{-z_o - r_s + \ell_p}{r_p}\right)\right)$$
$$+ \int_0^{\sinh^{-1}\left(\frac{-z_o+r_s-\ell_p}{r_p}\right)} dR'_{2D}(-z_o - \ell_p) + \int_0^{\sinh^{-1}\left(\frac{z_o+r_s-\ell_p}{r_p}\right)} dR'_{2D}(z_o - \ell_p)$$
$$+ \frac{1}{\sigma\pi\sqrt{r_p^2 - r_s^2}}\left(\tan^{-1}\left(\frac{\ell_p - z_o}{\sqrt{r_p^2 - r_s^2}}\right) + \tan^{-1}\left(\frac{\ell_p + z_o}{\sqrt{r_p^2 - r_s^2}}\right)\right) \tag{S99}$$

<u>Small sphere</u> $(r_s < \ell_p)$ <u>inside the pore</u> $(|z_o| < \ell_p - r_s)$:

$$R_b(z_o) = \frac{1}{2\sigma r_p} + \frac{2\ell_p - 2r_s}{\sigma \pi r_p^2} + \frac{2}{\sigma \pi \sqrt{r_p^2 - r_s^2}} \sin^{-1}\left(\frac{r_s}{r_p}\right) \tag{S100}$$

Interactive Desmos plot (for $z_o = 0$): https://www.desmos.com/calculator/zvan4euncv

## S11. Experimental Comparison Details

Figure 6a of the main article plots current-blockage data from Garaj *et al.*[2] who reported the measurement of DNA translocations through nanopores of different sizes fabricated in graphene membranes. The experiment were performed in 4M KCl solution, with a corresponding conductivity of $\sigma = 27.5\ S/m$, and under an applied voltage of 160 mV. To parametrize the conductance model, as per the Garaj report, pores were assumed to have a cylindrical geometry with a thickness of $L = 0.6\ nm$ and a diameter 0.2 nm smaller than the TEM-measured diameters, accounting for the non-conductive hydration layer. The DNA molecules were modelled as infinitely long cylinders with 2.2 nm diameter, as per Equation S72. Experimental $\Delta I$ values from Garaj *et al.* were estimated from Figure 3 of the original article, the values of which are reported in the following table:

| $d_{pore}$ (nm) | $\Delta I$ (nA) |
|---|---|
| 2.80 | 4.10 |
| 3.29 | 3.19 |
| 3.36 | 2.84 |
| 4.07 | 1.92 |
| 4.71 | 1.73 |
| 5.00 | 1.68 |
| 5.20 | 1.66 |
| 5.34 | 1.24 |
| 5.57 | 1.44 |
| 6.89 | 0.90 |

**Table S1.** Experimental data extracted from Garaj *et al.* used for Figure 6a.[2]

Figure 6b of the main article plots current-blockage data from Liu *et al.*[3] who reported the measurement of DNA translocations through nanopores of different sizes fabricated in MoS$_2$ membranes. The experiments were performed in 1M KCl solution, with a corresponding conductivity of $\sigma = 11\ S/m$. To parametrize the conductance model, pores were assumed to have a cylindrical geometry with a thickness of $L = 1.6\ nm$, as reported by Liu *et al.* Again, DNA molecules were modelled as infinitely long cylinders with 2.2 nm diameter, as per Equation S72. Note that for simplicity, no surface charge contributions were considered as opposed to the the original work. Experimental $\Delta G/G_o$ and $G_o$ values were estimated from Figure 5a of the original article, as reported in the following table:

| $G_o$ (nS) | $\Delta G/G_o$ |
|---|---|
| 24.04 | 0.340 |
| 30.34 | 0.126 |
| 38.29 | 0.156 |
| 38.11 | 0.061 |
| 43.50 | 0.114 |
| 43.69 | 0.050 |
| 49.72 | 0.104 |
| 54.20 | 0.0871 |
| 70.83 | 0.0657 |

**Table S2.** Experimental data extracted from Liu *et al.* used for Figure 6b.[3]

Figure 6c of the main article compares model predictions and experimental measurements of the fractional blockage $\Delta G/G_o$ for spherical nanoparticles passing through low aspect ratio pores. The articles used are from Davenport *et al*,[4] Bacri *et al*,[5] and Tsutsui *et al*.[6] Since these articles reported blockages in different manners, i.e. some giving current blockades and others fractional blockages, the following discussion and tables explain how the experimental $\Delta G/G_o$ values of Figure 6c were calculated from the blockade values reported in the different articles.

Davenport *et al.*[4] reported the fractional blockage values $\Delta G/G_o$ recorded for two different silica nanoparticles passing through silicon nitride membranes of various geometries in 0.1 M KCl solution. Experimental values were directly taken from Figure 2 of the main article and Figure S2 of the supporting information. Nanoparticles were modeled as spheres with diameters matching those of reported experiments. Nanopores geometries were assumed to be cylindrical with diameters and thicknesses matching those reported in the original article. Bigger discrepancy between experiments and model predictions are observed for thicker membranes, for which the cylindrical pore assumption differs most importantly from the gently tapered geometry expected of the pores, as discussed in the original report.

| $d_{pore}$ (nm) | $L_{pore}$ (nm) | $d_s$ (nm) | $\Delta G/G_o$ exp (%) | $\Delta G/G_o$ mod (%) | Error (%) |
|---|---|---|---|---|---|
| 260 | 50 | 57 | 0.76 | 0.74 | 4.0 |
| 260 | 50 | 101 | 2.57 | 3.39 | 31.9 |
| 260 | 100 | 57 | 1.15 | 0.62 | 46.1 |
| 260 | 100 | 101 | 4.95 | 3.67 | 26.0 |
| 280 | 500 | 57 | 0.87 | 0.226 | 74.1 |
| 280 | 500 | 101 | 2.74 | 1.34 | 51.1 |
| 328 | 50 | 57 | 0.67 | 0.374 | 44.2 |
| 328 | 50 | 101 | 2.31 | 1.69 | 26.6 |
| 234 | 100 | 57 | 1.43 | 0.827 | 42.1 |
| 234 | 100 | 101 | 5.37 | 4.95 | 7.9 |
| 307 | 500 | 57 | 0.81 | 0.181 | 77.6 |
| 307 | 500 | 101 | 2.81 | 1.07 | 62.1 |

**Table S3.** Experimental data extracted from Davenport *et al.* used for Figure 6c.[4]

Tsutsui *et al.*[6] reported the recordings of carboxylated polystyrene nanobead passing through $Si_3N_4$ membranes. The first experimental data point is the fractional blockage $\Delta G/G_o$ estimated from Figure 4b of the main article, whereas the rest of the data corresponds to the current blockages $\Delta I$ displayed in Figure S5 of the supporting information, recorded under a bias of 100 mV in 0.1 × PBS buffer. Experimental fractional blockage values $\Delta G/G_o = \Delta I/I_o$ were obtained by estimating the open pore current $I_o = \left(4L_{pore}/\sigma\pi d_{pore}^2 + 1/\sigma d_{pore}\right)^{-1}\Delta V$, where the solution conductivity was assumed to be 0.1 S/m. Nanoparticles were modeled as spheres with diameters matching those of reported experiments and nanopores geometries were assumed to be cylindrical with diameters and thicknesses matching those reported in the original article.

| $d_{pore}$ (nm) | $L_{pore}$ (nm) | $d_s$ (nm) | $\Delta I$ (nA) | $\Delta G/G_o$ exp (%) | $\Delta G/G_o$ mod (%) | Error (%) |
|---|---|---|---|---|---|---|
| 260 | 300 | 200 | - | 6.70 | 15.1 | 126.0 |
| 260 | 1200 | 900 | 1.519 | 13.3 | 15.2 | 14.2 |
| 260 | 1200 | 780 | 1.059 | 9.29 | 9.21 | 0.9 |
| 260 | 1200 | 510 | 0.2996 | 2.63 | 2.49 | 5.3 |

Table S4. Experimental data extracted from Tsutsui *et al.* used for Figure 6c.[6]

Bacri *et al.*[5] reported the recordings of silica nanoparticles in a 140 x 175 nm nanopore in a silicon nitride membrane of 50 nm thickness. The experimental value was taken from Figure 3c of the main article. Again, nanoparticles were modeled as spheres and nanopores as cylinders. In the original work, the nanopore is reported as having a diameter of 160 nm however, as noted, the nanopore is oblate and instead has dimensions of 140 x 175 nm. Interestingly, a 140 nm diameter used for the model results in a better agreement with experiments. Figure 6c plots the 160 nm comparison since that is the original reported value.

| $d_{pore}$ (nm) | $L_{pore}$ (nm) | $d_s$ (nm) | $\Delta G/G_o$ exp (%) | $\Delta G/G_o$ mod (%) | Error (%) |
|---|---|---|---|---|---|
| 160 | 50 | 85 | 13 | 9.16 | 29.5 |
| 140 | 50 | 85 | 13 | 13.88 | 6.8 |

Table S5. Experimental data extracted from Bacri *et al.* used for Figure 6c.[5]

Dutt *et al.*[7] reported the blockades of Bovine Serum Albumin (BSA) proteins in a 14.1 nm diameter pore in a 5.4 nm thick silicon nitride membrane. The experimental conductance blockage $\Delta G = 9.5\ nS$ was extracted from Figure 4f, whereas the open pore current of $I_o = 43\ nA$ under a 0.5V applied voltage from Figure 4e, such that $\Delta G/G_o = 9.5/43 \times 0.5 = 0.1105$. Although BSA is better represented as a prolate spheroid, a simplified analysis is presented in this work wherein BSA is interpreted as a sphere with a hydrodynamic radius of 3.48 nm.[8]

Saharia *et al.*[9] reported the blockade of human serum transferrin (hSTf) proteins through a 8.8nm diameter pore in a 1 nm thick hexagonal boron nitride (h-BN) membrane (Cited in Figure S1). The experimental conductance blockage was cited to be $\Delta G = 32.6$ nS whereas the open pore conductance $G_o = 19nS$ was calculated by extracting the open pore currents of Figure a-e and calculating the slope of the corresponding I-V curve. The protein diameter used for calculations was taken directly from the one cite din the article.

| Protein | $d_{pore}$ (nm) | $L_{pore}$ (nm) | $d_s$ (nm) | $\Delta G/G_o$ exp (%) | $\Delta G/G_o$ mod (%) | Error (%) |
|---|---|---|---|---|---|---|
| BSA | 5.4 | 5.4 | 6.96 | 11.1 | 9.16 | 31.6 |
| HS | 4.5 | 1 | 6.5 | 16.84 | 18.9 | 12.4 |

**Table S6.** Experimental data extracted from Dutt *et al.* and Saharia *et al.* used for Fig. 6c.[7,9]

We now show experimental nanoparticle translocation data from two different articles that were in strong disagreement with model predictions, and that were not included in Figure 6c. As per a 2018 report from Tsutsui *et al*,[6] the disagreement most likely arises from limited the temporal resolution altering the pulse amplitude because of the RC response arising from the pore resistance and membrane capacitance, which justified labeling these older data points as outliers and not including them in Figure 6c.

In 2016, Tsutsui *et al.*[10] reported the recordings of carboxylated polystyrene beads through nanopores in SiN membranes. Experimental values used for Figure 6c were calculated from the current blockages $\Delta I$ displayed in Figure 2 of the main text, which were recorded under a bias of 100 mV in 1× PBS buffer with a conductivity of $\sigma = 0.12821 \frac{S}{m} = \frac{1}{7.8\Omega}$, as cited in the original text. Experimental fractional blockage values $\Delta G/G_o = \Delta I/I_o$ were obtained by estimating the open pore current $I_o = \left(4L_{pore}/\sigma \pi d_{pore}^2 + 1/\sigma d_{pore}\right)^{-1} \Delta V$.

| $d_{pore}$ (nm) | $L_{pore}$ (nm) | $d_s$ (nm) | $\Delta I$ (nA) | $\Delta G/G_o$ exp (%) | $\Delta G/G_o$ model (%) | Error (%) |
|---|---|---|---|---|---|---|
| 1200 | 50 | 510 | 0.0880 | 0.603 | 2.49 | 313.2 |
| 1200 | 50 | 780 | 0.415 | 2.84 | 9.21 | 224.0 |
| 1200 | 50 | 900 | 0.601 | 4.12 | 15.2 | 270.0 |
| 1260 | 550 | 510 | 0.259 | 2.49 | 4.00 | 60.6 |
| 1260 | 550 | 780 | 0.719 | 6.92 | 14.9 | 115.9 |
| 1260 | 550 | 900 | 1.22 | 11.8 | 23.5 | 99.4 |
| 970 | 1050 | 510 | 0.351 | 6.70 | 6.41 | 4.4 |
| 970 | 1050 | 780 | 1.76 | 33.6 | 29.6 | 12.0 |
| 970 | 1050 | 900 | 2.28 | 43.6 | 55.3 | 26.9 |

**Table S7.** Experimental data extracted from Tsutsui *et al.*[10]

In 2012, Tsutsui *et al.*[11] reported the recordings of carboxylated polystyrene beads through nanopores in SiN membranes. Experimental values used for Figure 6c were calculated from the current blockages $\Delta I$ displayed in Figure 2 c and e and Figure 4a of the main text, which were recorded under a bias of 100 mV and 200 mV respectively in TE buffer with a conductivity of $\sigma = 0.0875$ S/m, as cited in the original text. Experimental fractional blockage values $\Delta G/G_o = \Delta I/I_o$ were obtained by estimating the open pore current $I_o = \left(4L_{pore}/\sigma \pi d_{pore}^2 + 1/\sigma d_{pore}\right)^{-1} \Delta V$.

| $d_{pore}$ (nm) | $L_{pore}$ (nm) | $d_s$ (nm) | $\Delta I$ (nA) | $\Delta G/G_o$ exp (%) | $\Delta G/G_o$ mod (%) | Error (%) |
|---|---|---|---|---|---|---|
| 1200 | 50 | 780 | 0.260 | 1.82 | 9.21 | 406.1 |
| 1500 | 400 | 780 | 3.92 | 3.82 | 8.33 | 118.0 |
| 1200 | 50 | 780 | 9.25 | 32.4 | 9.21 | 71.6 |
| 1200 | 50 | 900 | 9.03 | 31.6 | 15.2 | 51.8 |

**Table S8.** Experimental data extracted from Tsutsui *et al.*[11]